\documentclass[fleqn,usenatbib]{mnras}
\usepackage{newtxtext,newtxmath}
\usepackage[T1]{fontenc}
\usepackage{xspace}

\DeclareRobustCommand{\VAN}[3]{#2}
\let\VANthebibliography\thebibliography
\def\thebibliography{\DeclareRobustCommand{\VAN}[3]{##3}\VANthebibliography}

\usepackage{graphicx}	
\usepackage{amsmath}	
\usepackage{comment}

\def\Sref#1{Sect.~\ref{#1}\xspace}
\def\Fref#1{Fig.~\ref{#1}\xspace}
\def\Tref#1{Table~\ref{#1}\xspace}

\title[SLICK: Searching for strongly lensed GW]{SLICK: Strong Lensing Identification of Candidates Kindred in gravitational wave data} 

\author[S. Magare et al.]{
Sourabh Magare,$^{1}$\thanks{E-mail: sourabh.magare@gmail.com}
Anupreeta More,$^{1,2}$
Sunil Choudary$^{3}$
\\
$^{1}$The Inter-University Centre for Astronomy and Astrophysics, Post Bag 4, Ganeshkhind, Pune 411007, India\\
$^{2}$Kavli Institute for the Physics and Mathematics of the Universe (IPMU), 5-1-5 Kashiwanoha, Kashiwa-shi, Chiba 277-8583, Japan\\
$^{3}$The University of Western Australia (M013), 35 Stirling Highway, 6009 Perth, Australia
}

\date{Accepted XXX. Received YYY; in original form ZZZ}

\pubyear{2024}

\begin{document}
\label{firstpage}
\pagerange{\pageref{firstpage}--\pageref{lastpage}}
\maketitle

\begin{abstract}
By the end of the next decade, we hope to have detected strongly lensed gravitational waves by galaxies or clusters. Although there exist optimal methods for identifying lensed signal, it is shown that machine learning (ML) algorithms can give comparable performance but are orders of magnitude faster than non-ML methods. We present the \texttt{SLICK} pipeline which comprises a parallel network based on deep learning. We analyse the Q-transform maps (QT maps) and the Sine-Gaussian maps (SGP-maps) generated for the binary black hole signals injected in Gaussian as well as real noise. We compare our network performance with the previous work and find that the efficiency of our model is higher by a factor of 5 at a false positive rate of 0.001. Further, we show that including SGP maps with QT maps data results in a better performance than analysing QT maps alone. When combined with sky localisation constraints, we hope to get unprecedented accuracy in the predictions than previously possible. We also evaluate our model on the real events detected by the LIGO--Virgo collaboration and find that our network correctly classifies all of them, consistent with non-detection of lensing. 
\end{abstract}




\section{Introduction}
Gravitational waves (GWs) are solutions to Einstein's field equation in the linearized weak field limit: transverse waves of spatial strain caused by the time variation of the mass quadrupole moment of the source \citep[e.g.][]{schutz_2009}. With the first GW detection of GW150914, by the LIGO detectors at Livingston and Hanford \citep{PhysRevLett.116.061102}, the era of gravitational wave astronomy began. By the end of the third observing run (O3), over 90 events have been detected \citep[e.g.][]{PhysRevX.9.031040,PhysRevX.11.021053, PhysRevX.13.041039}. This also includes GW170817, which was the first coincident detection of GWs and electromagnetic observations from a binary neutron star merger \citep{PhysRevLett.119.161101, Abbott_2017}. With these detections came an avalanche of discoveries in fundamental physics like the speed of GWs \citep[e.g.][]{Abbott_2017b}, constraints on the mass of graviton \citep[e.g.][]{PhysRevD.100.104036}, constraints on the violations of lorentz invariance \citep[e.g.][]{PhysRevD.97.124023, PhysRevD.97.084040}, on violations of equivalence principle \citep[e.g.][]{Wei_2017} and many more \citep[e.g.][]{2019Natur.568..469M}. As the fourth observing run (O4) is ongoing and with the planned future observing runs like O5, Voyager and 3G detectors \citep[e.g.][]{Reitze2019Cosmic, Maggiore_2020et}, which will make the GW detectors even more sensitive, we can expect a flurry of interesting discoveries by the end of next decade.

One of the most anticipated discovery is of observation of gravitationally lensed GWs \citep[e.g.][]{Abbott_2021_lnsing}. GWs can get lensed if any intervening matter distribution like a galaxy or a cluster is present between the source and the observer \citep[e.g.][]{gunn1967, Wang1996}. Gravitational lensing of light is usually described in terms of a simplified framework called the geometric optics where light is treated as rays. However, in case of lensing of GWs, this does not hold true because the wavelength of GW may become comparable to the schwarzschild radius of the lens \citep[e.g.][]{1986ApJ...307...30D,PhysRevLett.80.1138,Takahashi_2003}. Thus, the wave-like effects- interference and diffraction- can come into play and one needs to make use of wave optics \citep[e.g.][]{schneider1992}. 
When we observe multiple images of the source because of the deflection of light by a massive foreground galaxy or cluster, it is called strong lensing. Just like EM waves, GWs can also get strongly lensed. As the GW signals from the mergers of compact binaries are transient, if they are strongly lensed we will observe identical signals separated by a time delay. In this work, we focus on detection of strong lensing of GW signals.

Till date, only gravitational lensing of EM sources has been observed \citep[e.g.][]{Dai2020Morse,Hannuksela2019,singer2019ligovirgo,theligoscientificcollaboration2023search,10.1093/mnras/stad2909}. Thus, the detection of lensing of GWs in itself can be a testable prediction of general relativity. Currently, gravitational lensing of EM has become a fundamental tool for astrophysics and cosmology. Lensing of GWs can present new avenue to constraint population of galaxies, dark matter distribution, precision cosmography  \citep[e.g.][]{10.1111/j.1365-2966.2011.18895.x, Li_2019, PhysRevLett.130.261401,Caliskan:2023zqm,Caliskan:2022hbu}, precise test for speed of gravitons \citep[e.g.][]{PhysRevLett.118.091102}, tests for gravitational wave polarizations \citep{PhysRevD.103.024038} and detecting intermediate and primordial black hole \citep[e.g.][]{PhysRevD.98.083005,Oguri_2020,Basak_2022}. Particularly, strong lensing can produce multiple images of gravitational wave signal which are separated by a time delay of minutes to weeks for galaxy. Rapid follow-up of the EM counterpart and the respective constraints can be used to give an early warning of the subsequent counterpart GW signal \citep{Magare_2023}. The rates of lensed BNSs and NSBHs, \citep[e.g.][]{10.1093/mnras/sty411,PhysRevD.97.023012,10.1093/mnras/sty2145,Magare_2023}, which assume that binary black hole traces star formation rate density, suggest confident detection of strongly lensed gravitational waves within this decade. In future observing runs, we expect to detect very large number of BBH events --$\mathcal{O}(10^{6})$ in 3G detectors \citep[e.g.][]{Abbott_2021_pop, Xu_2022}, a small fraction of them will be lensed pairs. It is important to identify the lensed pairs without loosing accuracy. Efficient, accurate and rapid detection methods are thus needed in the near future. 

Neural networks are steadily gaining popularity in the gravitational wave (GW) community. Several research groups have tested the reliability and accuracy of deep learning approaches on glitch classification, gravitational wave detection and parameter estimation \citep[e.g.][]{PhysRevD.97.101501,bahaadini2017deep,Shen_2019,Cavaglia_2019,PhysRevD.104.124057}. Some recent studies have demonstrated that CNNs could be used for identifying gravitational lenses using images of the lensed galaxies \citep[e.g.][]{jaelani2023survey, 10.1093/mnras/stx1492, Jacobs_2019, 2021ApJ...923...16L, 2023ApJ...954...68Z}. The real advantage of using a machine learning algorithm is that the only computationally expensive task is the training period of the neural network and is to be done only once. The evaluation time of any neural network on a sample data tends to be orders of magnitude faster than any other technique \citep{PhysRevD.94.044031}. We believe that the speed of machine learning algorithms can be of great utility for identifying strongly lensed pairs of GW signals provided accuracy is not compromised. The multiple copies of GW signal are separated by the time-delay and can have different amplitudes, however their phase evolution is expected to be identical for non-spinning circular binaries \citep[e.g.][]{Haris2018, Dai2017}. Thus, one way to infer, whether the given pair of signals are lensed or not, is by comparing the inferred posteriors on the intrinsic parameters. Strongly lensed pairs will show large overlap while unlensed signals will not show any significant overlap. The way to quantify this comparison is using bayesian model selection \citep[e.g.][]{Haris2018,Hannuksela2019,PhysRevD.104.124057,Janquart:2022zdd,Caliskan:2022wbh,cheung2023mitigating}. Given a pair of signal, calculate the bayes factor which is defined as the ratio of evidences of joint data of both signals. However, this is time consuming and computationally expensive. Bayesian parameter estimation of a BBH signal can take several hours to days. The situation gets worse, for example, we expect $\mathcal{O}(100)$ BBH events in O4 observing run and this number increases greatly for future observing runs \citep[e.g.][]{Abbott_2021_pop}. Like for O5 and voyager, number of BBH events is estimated to be of $\mathcal{O}(10^{3})-\mathcal{O}(10^{4})$, while for the 3G observing run, this number will likely be $\mathcal{O}(10^{6})$, out of which only a small percent could be strongly lensed \citep[e.g.][]{Xu_2022}. Thus, bayesian model selection technique may not be as efficient and fast for growing number of events. There are alternative pipelines e.g. \texttt{GOLUM} \citep{janquart2022golum}, \texttt{HANABI}  \citep{Lo_2023} and \texttt{phazap} \citep{Ezquiaga:2023xfe}. \texttt{GOLUM} is a faster technique to estimate joint parameters for a pair of events. \texttt{HANABI} takes in consideration the astrophysical information and selection effects, and uses hierarchical analysis for efficient computation of probabilities. \texttt{phazap} exploit the fact that phases are the best measured GW quantities and construct a joint distance which is a measure confidence level of lensing hypothesis for a pair of events. However, in terms of computational speed, these algorithms may take $\mathcal{O}(30)\rm ~minutes-\mathcal{O}(hours)$ for an event pair. Lower latency and complementary techniques are still desirable if we are expecting $\mathcal{O}(10^{3})-\mathcal{O}(10^{4})$ events which machine learning algorithms could facilitate.


\citet[][henceforth referred to as G21]{PhysRevD.104.124057} applied deep leaning algorithm on Q-transform maps \citep[QT maps,][]{SChatterji_2004} and sky localisation maps \citep[skymaps,][]{PhysRevD.93.024013} for identifying strongly lensed pairs. Their injections are added in gaussian noise. Further, the data is prepared such that a pair of (un)lensed events are superposed. Since random pair of events will have distinct projections in QT whereas lensing implies similar intrinsic parameters, the superposed maps will show differences between the lensed and unlensed pairs. However, we find that such QT maps can result in many false positives i.e. unlensed pairs getting classified as lensed. This is possible when the signal-to-noise ratio (SNR) of one event is sufficiently contrasting to the SNR of another event in the pair. In this work, we attempt to bypass this issue by using QT maps along with the Sine Gaussian Maps  \citep[SGP-maps][]{PhysRevD.107.024030}. We find that using both representations increases the efficiency in identifying lensed pairs as compared to QT maps alone. Further, to improve the robustness of our network, we train with a dataset consisting of injections in real noise as well as in gaussian noise and compare their performance. 

The rest of the paper is organized as follows. In \Sref{sec:method}, we describe the type of input data used, sample preparation, suitability of data for binary classification, the neural network models that are developed for the Strong Lensing Identification of Candidates Kindred (\texttt{SLICK}) pipeline. \Sref{sec:res} describes our results on various datasets including the comparison with the previous work by G21. \Sref{sec:summ} gives the summary of our work.

\section{Methodology}
\label{sec:method}
The \texttt{SLICK} pipeline is designed to take simulated or real GW signals, generate input data in the form of images and run a custom neural network to predict whether a pair of signals are strongly lensed or not. We describe below the components of the \texttt{SLICK} pipeline along with the methodology.

\subsection{Input Data Representation}
\label{ssec:inpdata}
We use two type of data representations - QT Maps and SGP Maps. 

\begin{figure}
    \centering
    \includegraphics[width = \columnwidth]{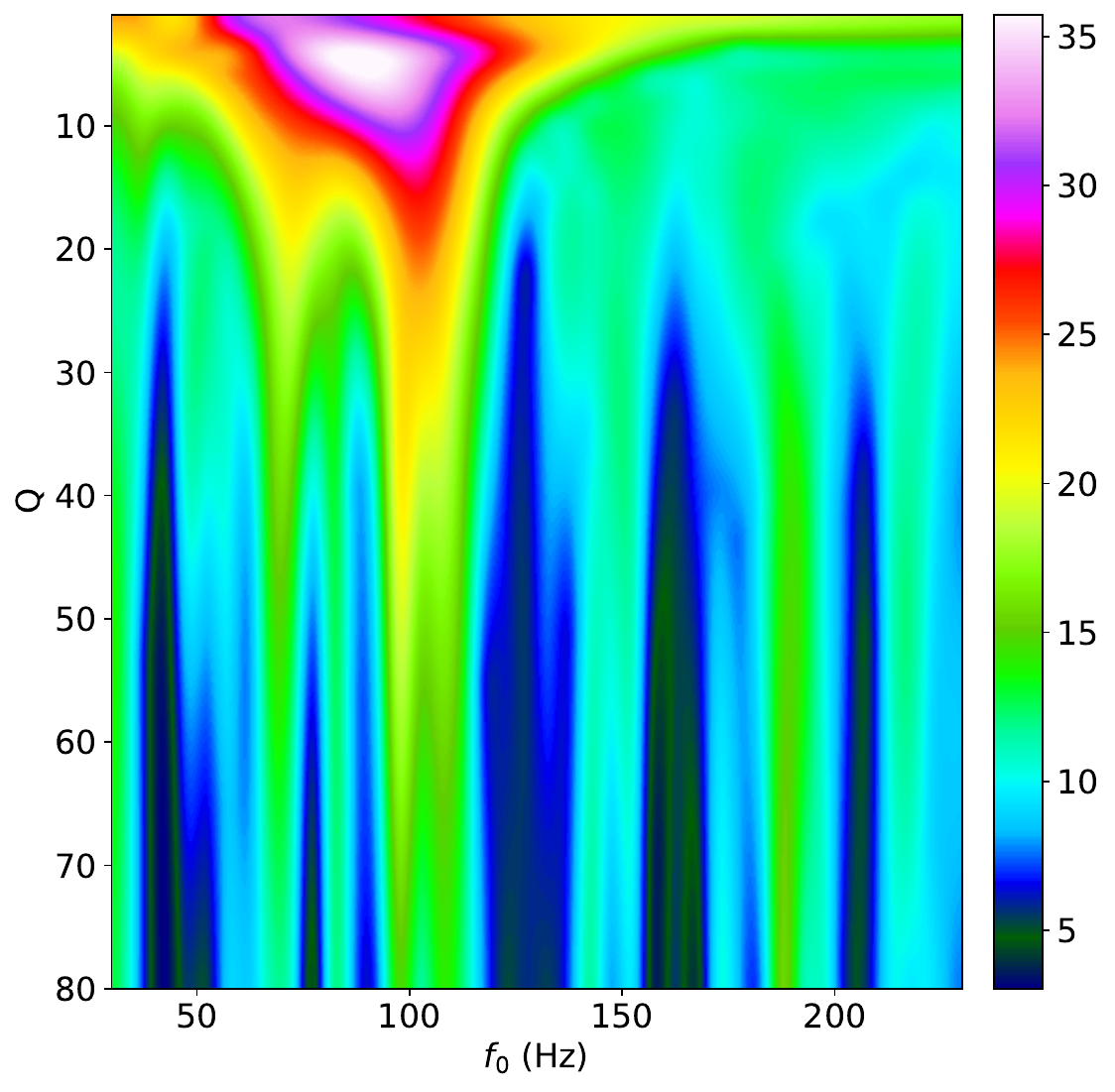}
    \includegraphics[width=\columnwidth]{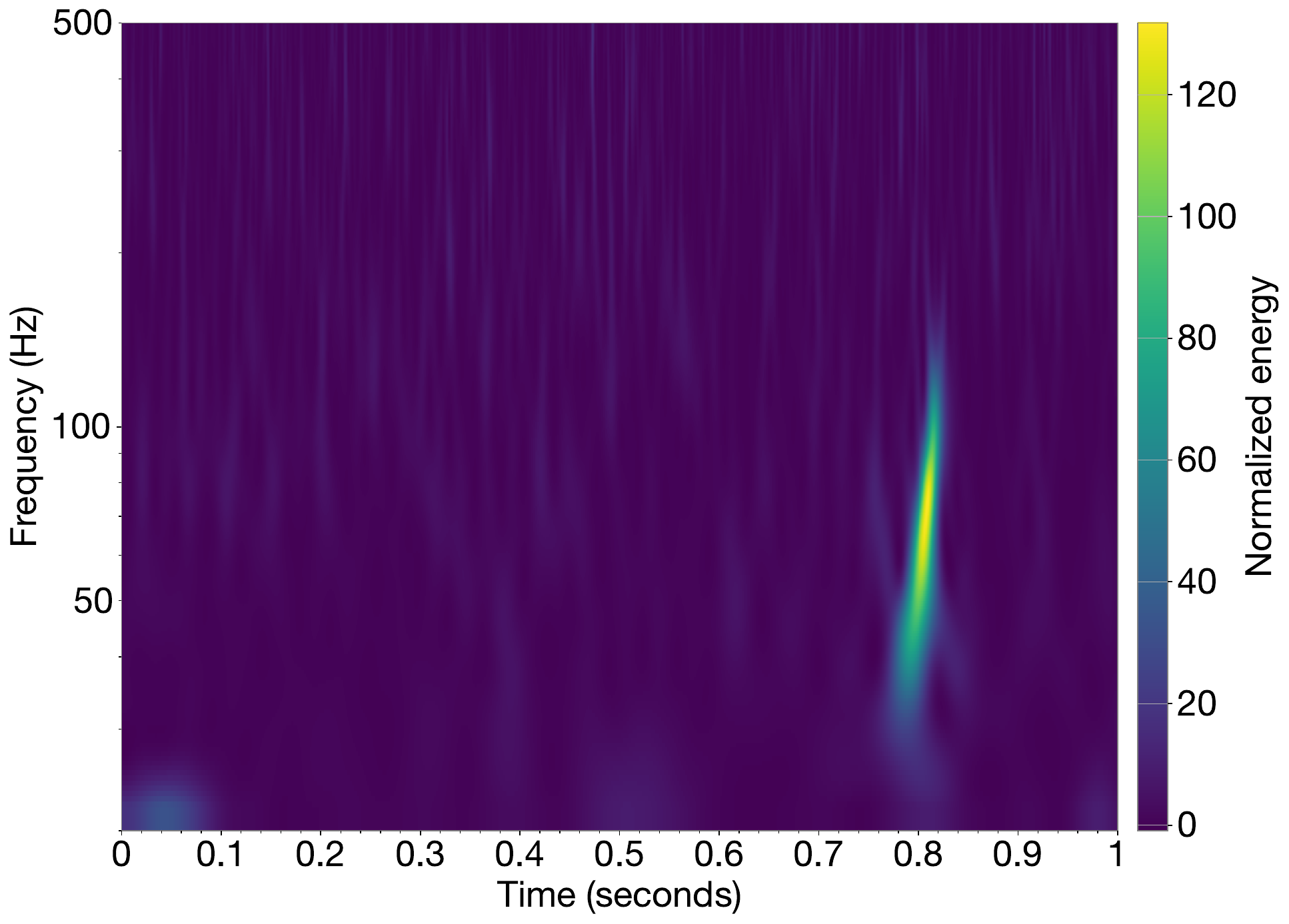}
    \caption{ {\it Top:} Sine-Gaussian projection (SGP) map of a simulated BBH signal injected in gaussian projected O4 noise. The color represents the value of the projection of the timeseries data on the sine-gaussian function. {\it Bottom:} Q-transform map of a BBH signal injected in gaussian noise. The color represents the normalized energy of the signal.}
    \label{fig:SGP_QT}
\end{figure}

The Q-transform \citep{SChatterji_2004} technique has long been used as a visualization tool in the LIGO--Virgo data analysis. It is a modified short-time fourier transform where the analysis window duration varies inversely with frequency such that the plane of time-frequency is covered by the tiles of constant $Q$. 
The continuous Q-transform is given by
\begin{equation}
    x(\tau, f) = \int_\infty^\infty x(t)\, w(t-\tau) \,\exp({-2 \pi i f t})\, {\rm d}t
\end{equation}
where  
$w(t-\tau,f)$ is a window function centered at $\tau$ and depends on the quality factor $Q$, 
\begin{equation}
    Q = \frac{f}{\delta f}
\end{equation}
The discrete version of the above is more appropriate for gravitational wave data, as this data is collected in discrete time bins, 
\begin{equation}
    x[m,k] = \sum_{n=0}^{N-1} x[n]w[n-m,k] \exp{(-2\pi i nk/N)} 
\end{equation}
The value of $Q$ is set in the following way: When we perform Q-transform of a signal, we have to specify a range of values of $Q$. The value of $Q$ which corresponds to the largest SNR is chosen.
This transformation creates a tile in time and frequency domain and the value in each tile corresponds to the energy of the signal.
The \Fref{fig:SGP_QT} shows the QT map (bottom) of a BBH merger event.

The other kind of data representation, we use, is called the SGP maps \citep{PhysRevD.107.024030}. 
It is mathematically expressed as 
\begin{equation}
    g(t) = A\exp{(-4\pi f_{0}^{2}}\frac{(t-t_{0})^{2}}{Q^{2}})\cos{(2\pi f_{0}t \phi_{0})}
\end{equation}
where $A$ is the constant ampitude, $Q$ is the quality factor, $f_{0}$ is the central frequency, $\phi_{0}$ is the phase and $t_{0}$ is the central time. Let $x(t)$ be the timeseries data, then the projection of timeseries data on the sine-Gaussian is given by 
\begin{equation}\label{sgp_projection_equation}
    (x, g) = 4 {\rm Re} \int_{f_{\rm{lower}}}^{f_{\rm{higher}}} \frac{\tilde{x}^{*}(f) \tilde{g}(f)}{S_{n}(f)} df
\end{equation}

where $S_{n}(f)$ is the power spectral density (PSD) of the noise and $\tilde{x}(f)$ and $\tilde{g}$(f) are the fourier transform of the $x(t)$ and $g(t)$, respectively.

For constructing the SGP maps, we evaluate the above projection in the space of $f_{0}$ and $Q$. 
The chosen range of $f_{0} \in [20,240]$~Hz and $Q \in [2, 80]$.  
The \Fref{fig:SGP_QT} shows the SGP map (top) of a BBH merger event.

From the QT and SGP maps of an event pair, we look for features that can help us discern a lensed pair from an unlensed pair. Usually the lensed events are a pair of events which have the same intrinsic parameters like mass and spins but with different amplitudes \citep[e.g.,][]{schneider1992,Takahashi_2003,dodelson2017}. This implies that the pair of lensed signals should have mostly similar projections in the QT as well as SGP maps modulo some differences from extrinsic parameter and arrival times. However, a random pair of unrelated signals are less likely to have similar intrinsic parameters and thus, their QT and SGP maps are not expected to be strongly similar. Such similarities will only rarely occur by chance.

\begin{figure*}
    \centering
    \includegraphics[scale=0.55]{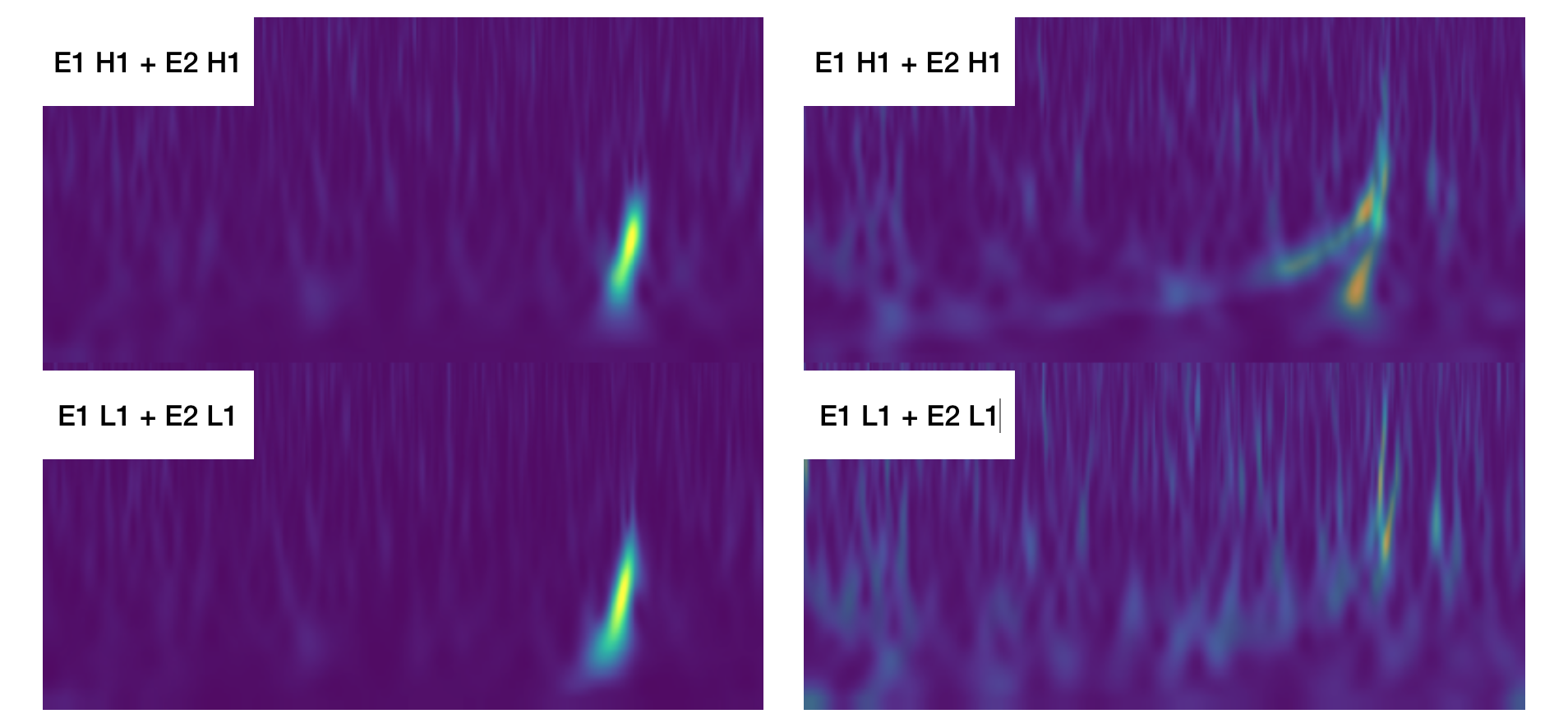}
    \caption{
    Sample format of input data to the neural network that analyses the QT maps. The QT maps for a pair of events (E1 and E2) are superposed and the QT maps from two detectors (H1 and L1) of the corresponding events are placed one below the other. {\it Left:} Sample QT maps for a lensed event pair. {\it Right pair:} Sample QT maps for an unlensed event pair. }
    \label{fig:QT_input_image}
\end{figure*}

\begin{figure*}
    \centering
    \includegraphics[scale=0.52]{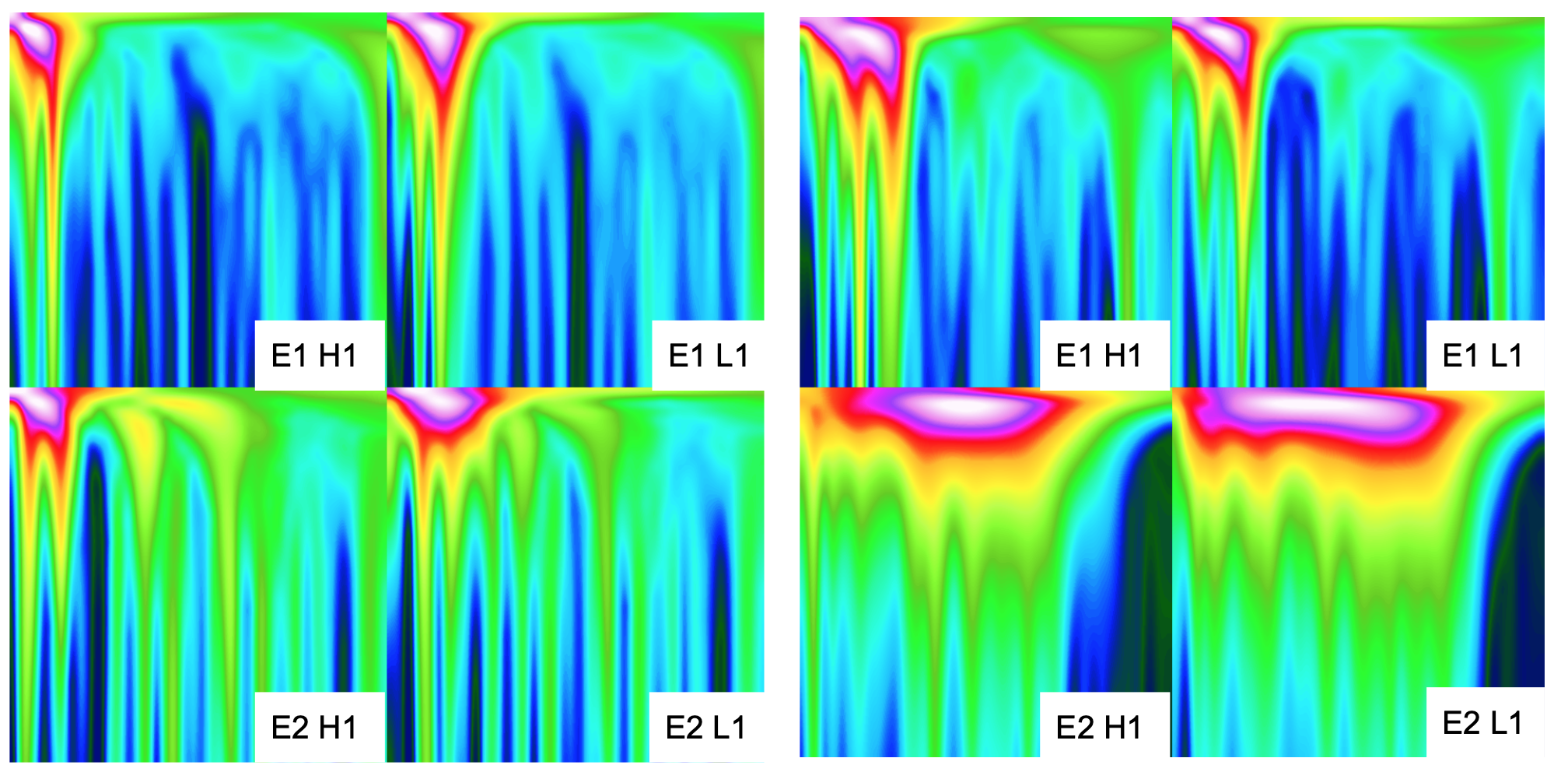}
    \caption{Sample format of input data  to the neural network that analyses SGP maps. The SGP maps are produced in a 2x2 grid format where the labels imply the following - E1H1: Event1 in H1 detector, E2H1: Event2 in H1 detector, E1L1: Event1 in L1 detector, E2L1: Event2 in L1 detector. {\it Left:} Sample SGP map for a lensed event pair. {\it Right:} Sample SGP map for an unlensed event pair. }
    \label{fig:SGP_input_image}
\end{figure*}

In \Fref{fig:QT_input_image} and \Fref{fig:SGP_input_image}, we show the input data representations of lensed and unlensed images. For the QT maps per event pair as input data, we choose a 2$\times$1 grid format where a row corresponds to the detector (H1 or L1) and each map is a superposition of the pair of events. It is clear from \Fref{fig:QT_input_image}, that the lensed events have similar projections and near perfect overlap while the maps of the unlensed event pairs do not have as much overlap. This serves as the primary distinguishing criterion between the lensed and unlensed event pairs. For the SGP maps of an event pair, we choose a $2\times2$ grid format instead. Here, each row corresponds to each event in the pair while each column corresponds to each detector, H1 and L1. It is clear from the \Fref{fig:SGP_input_image} that for the lensed pair, higher projections are seen in the same region of the $Q-f_0$ plane, while the unlensed event pair may not have similar projections. As before, these patterns help distinguish the lensed pair from the unlensed ones.

\subsection{Image classification problem}
\label{ssec:img_class}
We see from \Fref{fig:QT_input_image} that for the QT maps, superposed lensed pair of events show different morphological features than superposed unlensed pair of events. The lensed pair has different amplitudes but similar time evolution since their intrinsic parameters are the same, while the unlensed pair, in general, will have distinct intrinsic parameters. These make their morphologies distinct. Thus, the problem of identifying the lensed and unlensed pairs can be framed as a binary image classification problem, and can be well handled by Deep Learning networks \citep{SCHMIDHUBER201585}. 


Since the lensed pairs will have identical frequency evolution, we should expect that the projection of signal in the $Q-f_0$ plane of the SGP should be identical. Our expectations are validated from \Fref{fig:SGP_input_image}. The lensed pair is projected on the identical part of the $Q-f_0$ plane while a randomly selected unlensed pairs is projected on different parts of this plane. In case of SGP maps input data, we haven't superposed the pair of signals, but rather kept them as $2\times2$ tiles. We expect that the SGP maps provides the complementary information to QT maps and together with QT maps will help reduce the false positives.

\subsection{Generation of Sample }
\label{ssec:gen_samp}
We prepare the timeseries data by injecting a simulated GW signal in real and gaussian noise. We follow \cite{10.1093/mnras/stac1704} to generate lensed and unlensed populations of binary black holes (BBHs). We consider the mergers of BBHs as the background source and massive early type galaxies as the lens galaxies. The density profile of the galaxies is given by a Singular Isothermal Ellipsoid (SIE) lens model \citep{Kormann1994}. The SIE model produces either two lensed images (double) or four lensed images (quad). In case of a quad, we choose the brightest two images to define a lensed pair such that the fainter event in the pair has a network SNR $>8$. The same SNR condition is applied to a double.

We generate $\approx$17000 lensed pairs and $\approx$4000 unique unlensed events. We then divide this set into training, validation and testing samples. For training, we use 12000 lensed pairs and make random pairs from the catalog of unlensed events, to get 12000 unlensed pairs. Similarly, we prepare a sample of 2000 each of lensed and unlensed pairs for the validation sample. We use the remaining samples as the test sample. 
We use the IMRPhenomPv2 \citep[e.g.][]{PhysRevLett.113.151101,PhysRevD.93.044007} waveform approximant, implemented in LALSuite, to generate the simulated GW signal. We take into account the antenna patterns from the two LIGO detectors, Livingston (L1) and Hanford (H1), to produce the final ``observed" signal, before adding the noise, using the PyCBC package. While generating the lensed signals, we also apply the morse phase shift to the type II images (saddle points). The morse phase for both the doubles and typically, the two brightest images of the quad are $\frac{\pi}{2}$. Although, we do not expect that the phase shift effects will manifest in either the QT maps or the SGP maps. 

We train and test the neural networks on gaussian as well as real detector noise which are referred to as Dataset-G and Dataset-R, respectively (see \Tref{tab:datasets}). To generate the Dataset-G,  we use the aLIGO PSDs\footnote{https://dcc.ligo.org/LIGO-T2000012/public} to produce 16-sec-long $\approx$21~000 unique noise segments and inject our simulated signals in them. To generate the Dataset-R, we extract a 60-min data segment from the third observing run (O3) of LIGO-Virgo \citep{LIGOScientific:2019lzm, KAGRA:2023pio}. This serves as our parent noise data chunk from which again $\approx$21~000 unique noise segments are extracted with a random start-time. The simulated signals are then injected in these 16-sec-long real noise segments.   

Further, we also evaluate the performance of our neural network on real events detected in O2 and O3 called the Dataset-LHV \citep{PhysRevX.9.031040,Abbott_2021,PhysRevX.13.041039}. We take 8 events from O2: between GW170104 to GW170823, and 36 events from O3: between GW190408 to GW180930. We pair these events with each other to obtain a total of 1035 pairs. We assume that all of the real event pairs are unlensed and evaluate the neural network based on the mis-classification of these unlensed pairs as lensed.   

To generate the QT maps, we use the Q-transform method in GWPY package on the timeseries data. In the Q-transform method, we choose the {\it frange}$= [20, 512]$ and also set the {\it whiten} parameter to be true.
We chop the timeseries to a length of 1 second with the peak of the merging signal to be located at 0.8~s.  We set the frequency from 30~Hz to 1000~Hz and $q=$~(4, 64). The same timeseries is then used to generate the SGP maps. For generating the SGP maps, we choose the range of $f_{0} \in [20,240]$~Hz and $Q \in [2, 80]$. The data points in the chosen range are then sampled uniformly and for each data point, the time series is projected on the Sine Gaussian function using Eq. \ref{sgp_projection_equation}. This projection is then presented using a color map and spline interpolation is used for smoothing of pixels.

\begin{table}
\centering
\caption{Description of various datasets used in our analysis}
\label{tab:datasets}
\begin{tabular}{ |p{3cm}|p{5cm}|  }
\multicolumn{2}{|c|}{} \\
\hline
Dataset & Description \\
\hline
G & Simulated GW signals are injected in {\bf gaussian} noise which is generated using Advanced LIGO PSDs.  \\
R & Simulated GW signal are injected in {\bf real} noise. These noise segments are extracted from O3 run for H1 and L1 detectors.  \\
LHV & Detected events from O2:between GW170104 and GW170823 and O3: between GW190408 and GW180930. Total 46 events which gives 1035 pairs. \\
DST & BBH parameter samples used by G21. Using these samples, GW signals are simulated and injected in {\bf gaussian} noise which is generated using Advanced LIGO PSDs. \\
\hline
\end{tabular}
\end{table}

\begin{figure}
    \centering
    \includegraphics[width=\columnwidth]
    {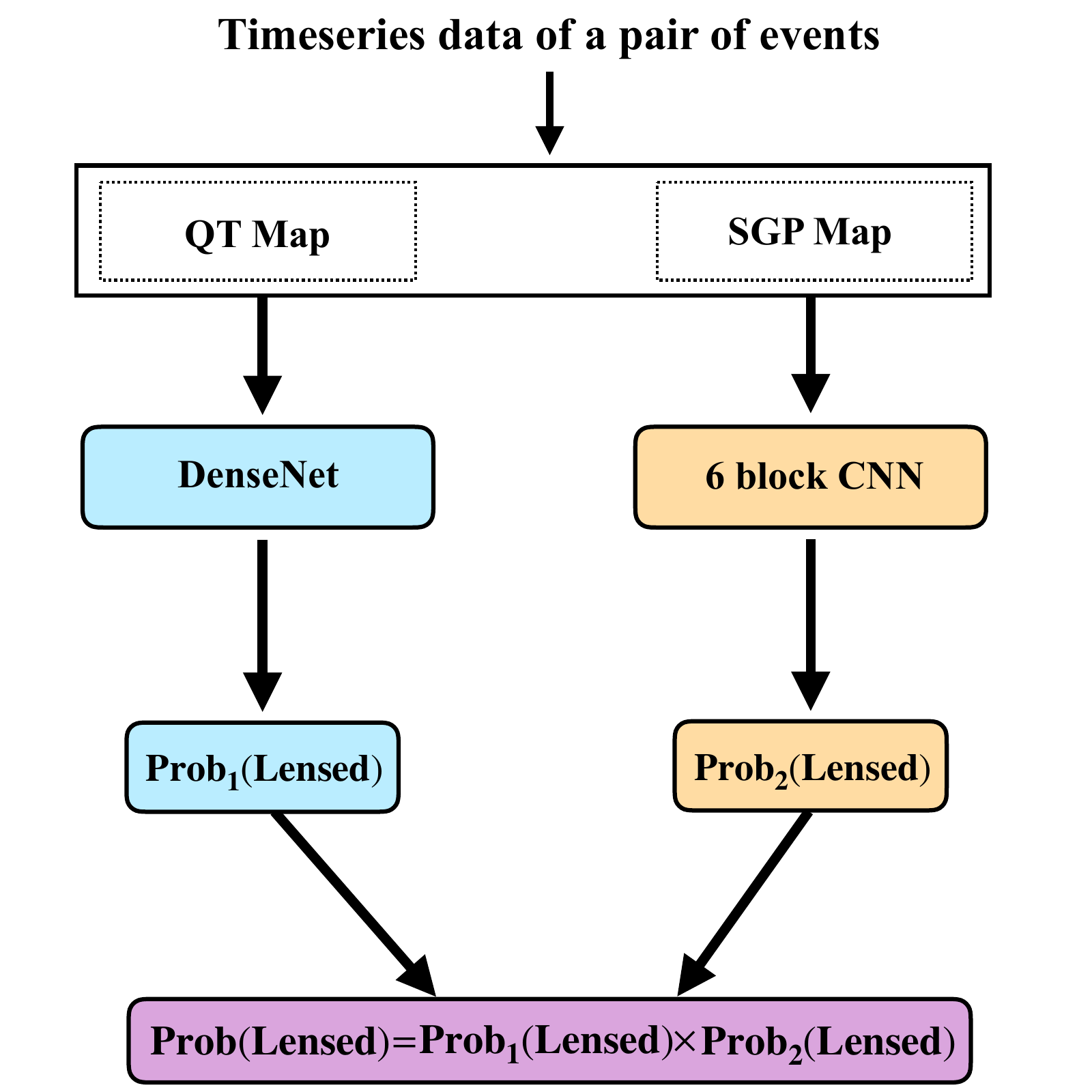}
    \caption{
    Visual representation of the workflow of \texttt{SLICK}. Using the timeseries data for a pair of events, QT and SGP maps are generated. These maps serve as input data through their respective neural network models where the output from each network is a probability for the pair of events to be lensed. The final output of our \texttt{SLICK} pipeline is a combined probability. 
    }
    \label{fig:workflow}
\end{figure}


\subsection{Deep Learning Models}
\label{ssec:mlmodel}
In this section, we give a brief description about the neural networks that are used in the \texttt{SLICK} pipeline. The readers may refer to \cite{deep_article} and \cite{SCHMIDHUBER201585} for a detailed review on deep learning.

A specialized class of deep neural network, which have been proven highly effective in computer vision task like image recognition are the Convolutional Neural Networks \citep[CNNs, e.g.][]{huang2018densely,ImgNetClass,deep_article}. CNNs typically have the following structure- Convolutional layers, Pooling layers and Fully connected layers. Convolutional layers perform the convolution operation-which involves the sliding a matrix called filter or kernel over the input to perform element-wise multiplication to generate output which is called a feature map or activation map. Pooling layers are also known as downsampling, which performs dimensionality reduction-i.e. to reduce the spatial dimensions of the feature map while also retaining relevant information. Max pooling and Average pooling are the two commonly used pooling layers. After many blocks of convolutional layers followed by pooling layers, a typical CNN ends with fully connected layers. Here, each node in a layer is connected to every node of the previous layer. These layers perform the classification based on the extracted features from the convolutional layers.
    
\begin{figure*}
    \centering
    \includegraphics[scale = 0.55]{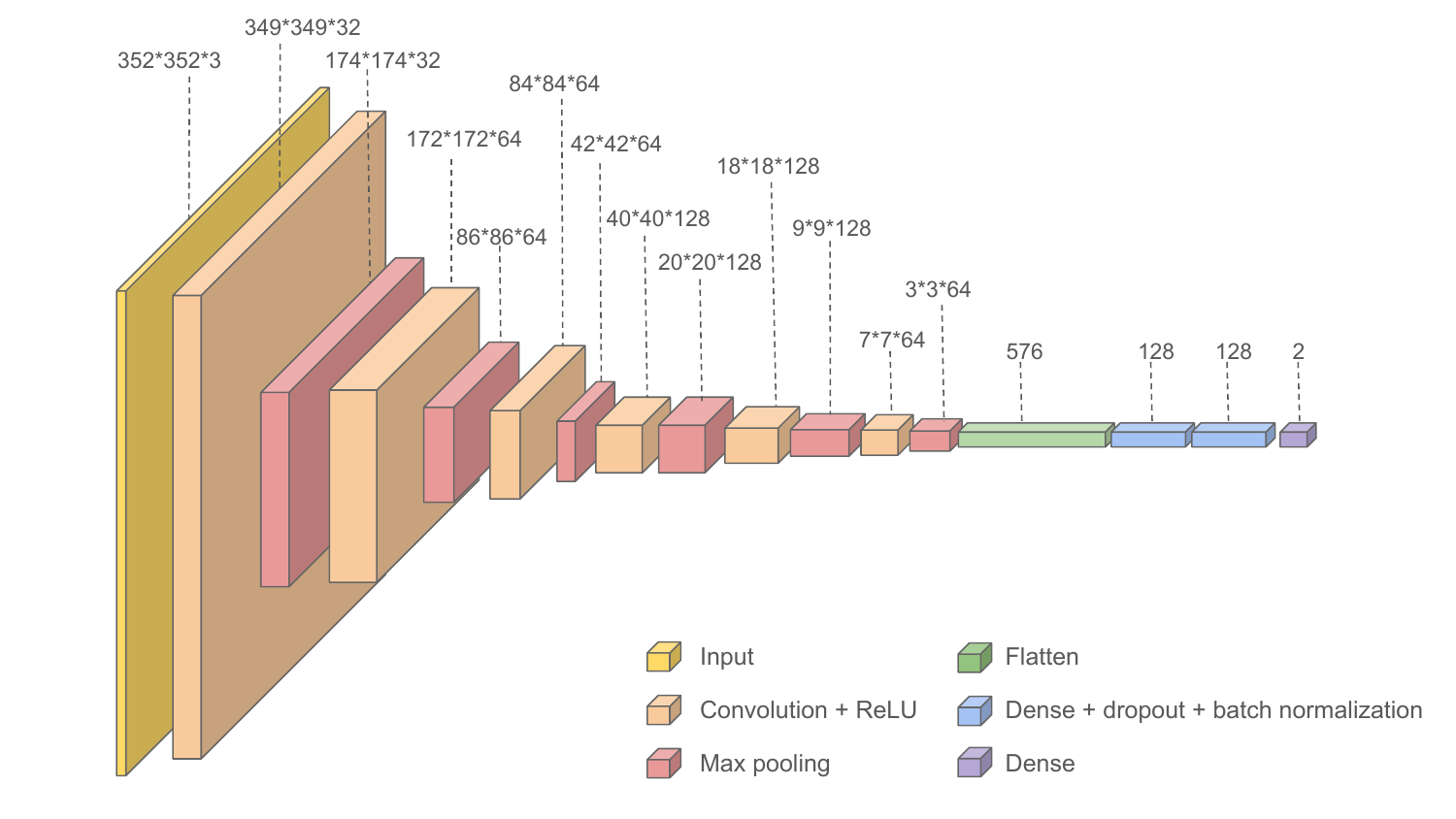}
    \caption{The 6 block CNN model used for classification of lensed and unlensed pairs with the SGP maps as input data.}
    \label{6_CNN}
\end{figure*}

We build two independent ML models- one that learns from QT maps and other learns from SGP maps for binary classification of lensed and unlensed pairs.
First network is the Densenet121 model that is trained on the superposed QT maps. The \Fref{fig:QT_input_image} shows the input image data given to this neural network. The second network is a standard CNN model which consists of 6 convolution blocks and is trained on the SGP maps. The \Fref{fig:SGP_input_image} shows the input data given to this network. The outputs of both these networks are combined to get the final probability that the given pair of events are lensed or unlensed. \Fref{fig:workflow} shows a graphical representation of the overall workflow.

{\it DenseNet121:} DenseNet (Densely Connected Convolutional Networks) are deep convolutional networks in which each layer receives direct inputs from all preceding layers-these are called dense connections \citep{huang2018densely}. Using this design they reduce the problem of vanishing gradients, enhance feature propagation and reduce the number of parameters. DenseNet121 refers to a specific architecture with 121 layers. However, it is still difficult to train this neural network from scratch, thus we make use of transfer learning. Transfer learning is a technique where a network which is trained for a particular task is reused for a different but related task \citep[e.g.][]{yosinski2014transferable}. For our binary classification problem of distinguishing between lensed and unlensed images, we use DenseNet121 model which is pre-trained of ImageNet dataset \citep{imgNetdata}. We make some changes to the network like adding new fully connected  layers along with Dropouts after the DenseNet121 model. We then set the first 10 layers to be non-trainable. Densenet121 is used on the QT maps to distinguish between lensed and unlensed events. We train this network on 12000 lensed and 12000 unlensed pair QT maps of {\it Dataset-R} as described in section 2.2.

{\it 6-CNN model:} We use a 6-CNN layered network on the SGP maps to distinguish between lensed and unlensed pair of events. \Fref{6_CNN} shows the detailed network architecture of this model. We train this network from scratch on 12000 lensed and 12000 unlensed pairs of SGP maps of two different datasets- {\it Dataset-G} and {\it Dataset-R}. 

After training DenseNet and CNN model individually, we combine output of both to get the final probability that the given pair of events are lensed or unlensed, as shown in the workflow \Fref{fig:workflow}. We name the networks with respect to their training dataset as shown in the \Tref{Network_nomenclature}.

\begin{table}\label{Network_nomenclature}
\centering
\caption{Various types of network models and datasets used therein.}\label{tab:model}
\begin{tabular}{ |p{1.7cm}|p{3cm}|p{2.5cm}  }
\multicolumn{3}{|c|}{} \\
\hline
Model & QT maps Training & SGP maps Training \\
\hline
Model-Q$_{\rm r}$ & Dataset-R & -  \\
Model-Q$_{\rm r}$S$_{\rm g}$ & Dataset-R & Dataset-G \\
Model-Q$_{\rm r}$S$_{\rm r}$ & Dataset-R & Dataset-R \\
\hline
\end{tabular}
\end{table}

\vspace{0.5cm}
The nomenclature adopted is a follows: for Model-Q$_{\rm \alpha}$S$_{\rm \beta}$, the QT maps part of the network ({\it DenseNet121}) is trained on dataset-$\alpha$ and the SGP part of the network ({\it 6-CNN}) is trained on dataset-$\beta$. For Model-Q$_{\rm r}$, there is no SGP-map part of the network, so `S'\, is ignored from its name.

\begin{figure}
    \centering
    \includegraphics[width=\columnwidth]{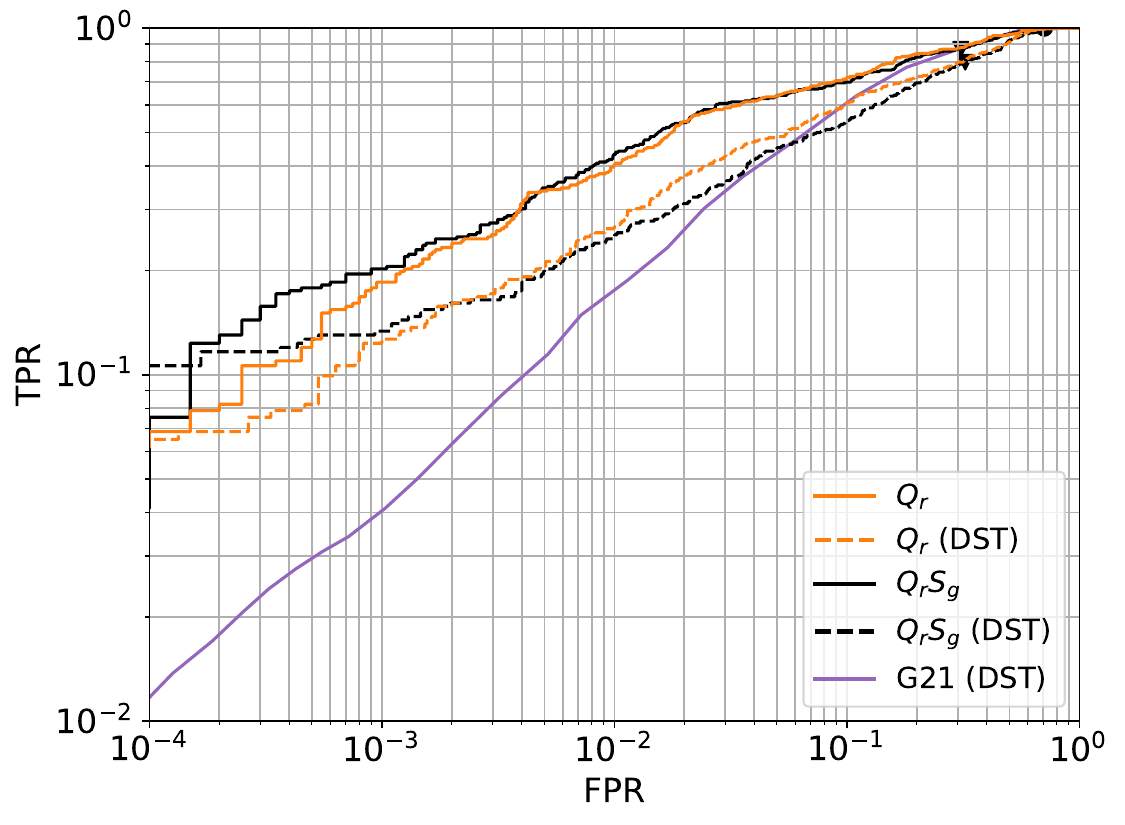}
    \caption{Evaluation of our models on Dataset DST and our datasets along with the comparison of results from G21. We choose 300 lensed pairs and 30,000 unlensed pairs from each of the dataset for a fair comparison. The Model-Q$_{\rm r}$S$_{\rm g}$ (solid black curve), comprising QT and SGP maps, performs well at low FPRs as compared to the Model -Q$_{\rm r}$ (orange curve), with QT only data, tested on both DST and our datasets.  }
    \label{haris-comparision}
\end{figure}

\section{Results and Discussion}
\label{sec:res}

\begin{figure*}
    \centering
    \includegraphics[scale=0.7]{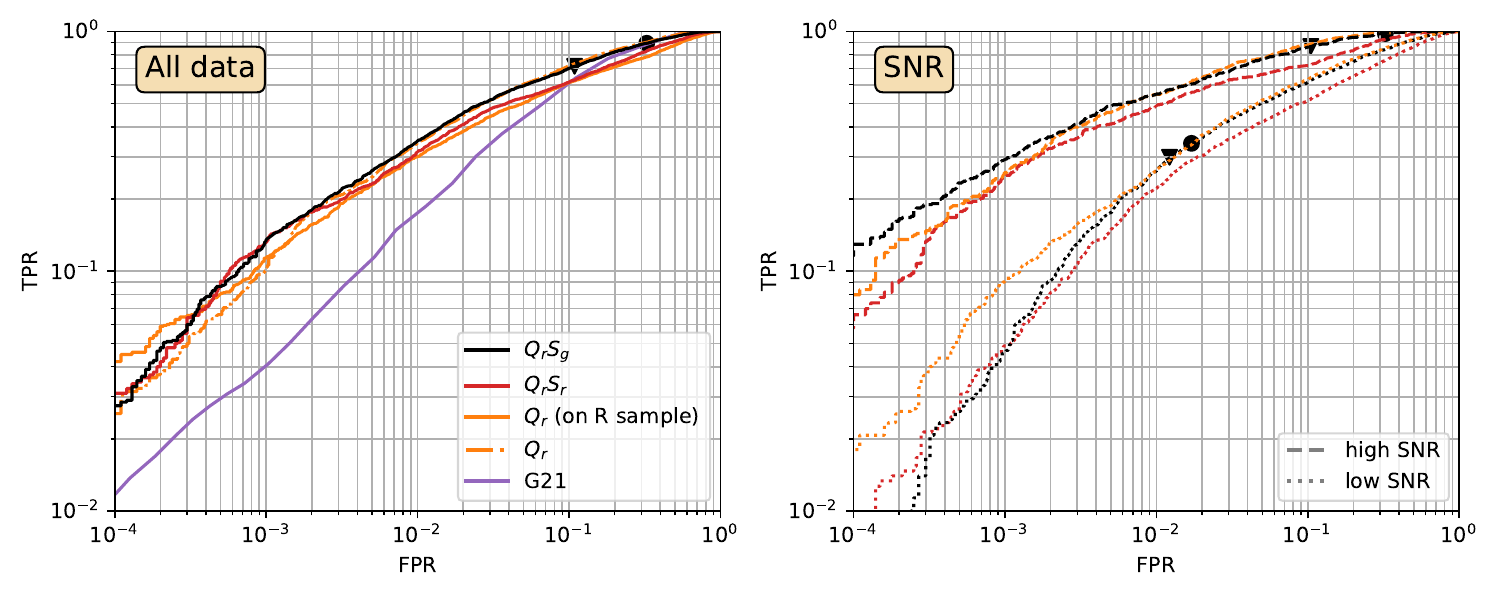}
    \includegraphics[scale=0.7]{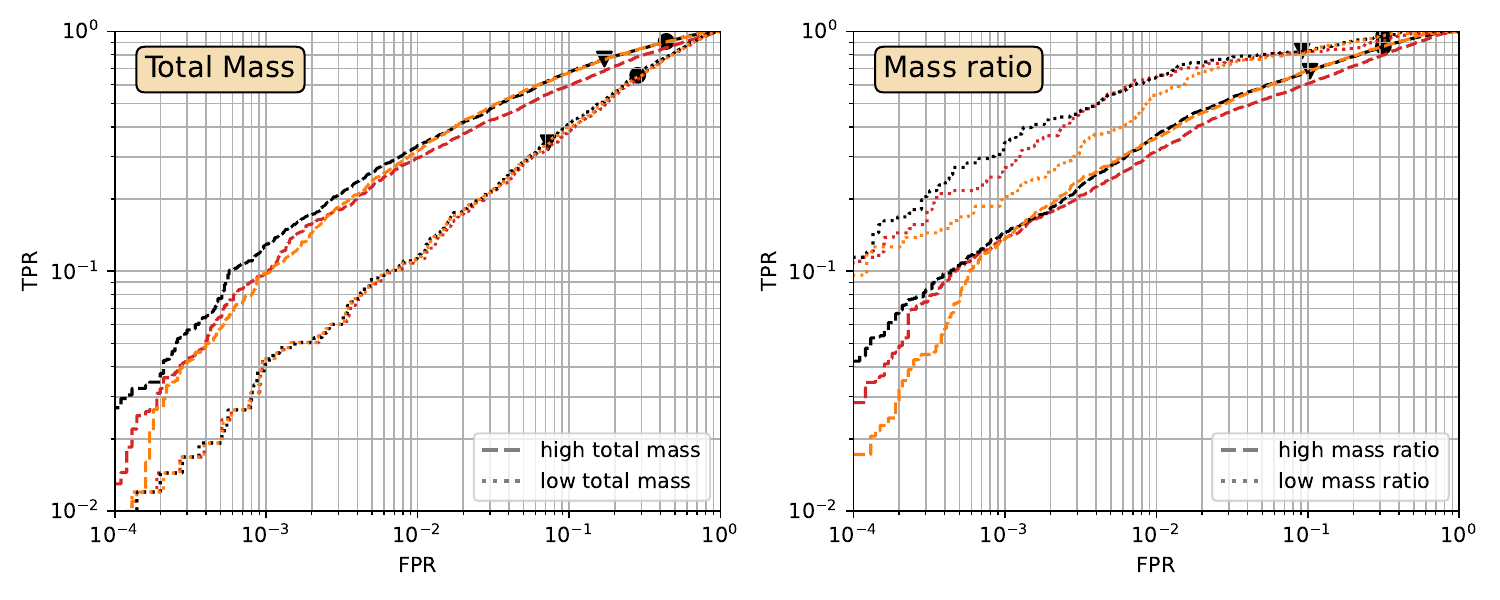}
    \caption{ Performance of our networks on test samples made from Dataset-G. {\it Top left panel:} The ``All Data" consists  of 2,000 lensed pairs and 100k unlensed pairs. Here, we find that the hybrid trained model (solid black curve) performs better than QT only model especially (solid red curve) at low FPRs. The Model-Q$_{\rm r}$ is tested on both Dataset-G (dashed-dotted orange) and Dataset-R (solid orange) with the former performing moderately better. We then create a subsample of the ``All Data" as a function of SNR ({\it top right panel}), total mass ({\it bottom left panel}) and mass ratio ({\it bottom right panel}). These subsamples are then divided in high and low bins (e.g. high and low snr). Here, the {\it dashed} correspond to ROCs at high values (e.g. high snr) and the {\it dotted} correspond to ROCs at low values (e.g. low snr). The {\it black triangle} and {\it black circle} correspond to 0.9 and 0.7 probability, respectively.}
    \label{fig:all_MQS}
\end{figure*}

We assess the performance of the trained models by evaluating them on various test datasets. We use the Receiver Operating Characteristic (ROC) curves as a statistic of comparison between different trained models. The ROC plot describes a trade off between the true positive rate (TPR, y-axis) and false positive rate (FPR, x-axis). The TPR is the ratio between the correctly classified lensed events to the total number of lensed events, while the FPR is the ratio between incorrectly classified unlensed events to the total number of unlensed events. The ideal ML model should have a TPR and a FPR as 1 and 0, respectively, at all thresholds.

Firstly, we compare the performance of our ML models with the work of G21. We use the DST dataset on which the G21 network is evaluated. For convenience, we work on a subsample of DST dataset that includes 300 pairs of lensed events and 30,000 pairs of unlensed events. We further test our model on a sample of randomly selected lensed and unlensed pairs with the same ratio from our Dataset-G (see \Fref{haris-comparision}). 
For comparison, we also plot the ROC curve of G21, which is evaluated on the QT maps of the Dataset-DST. We see that both the Model-Q$_{\rm r}$ and Q$_{\rm r}$S$_{\rm g}$ are performing better than G21 at lower FPRs ($<10^{-3}$). Further Model-Q$_{\rm r}$S$_{\rm g}$, trained on both the QT maps and the SGP maps, performs better compared to the Model-Q$_{\rm r}$, trained only on the QT maps. 

\begin{figure}
    \centering
    \includegraphics[width=\columnwidth]
    {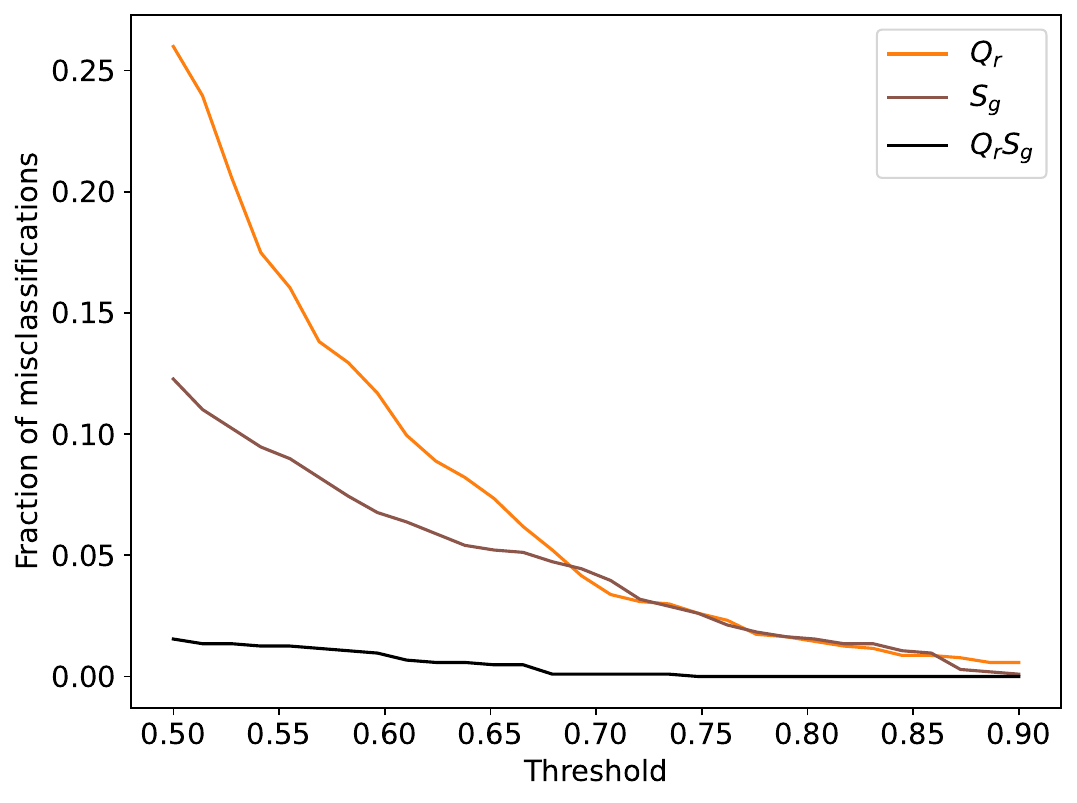}
    \caption{Fraction of misclassifications of Dataset LHV as a function of network threshold. Evaluation is done on a total of 46 events resulting in 1035 pairs where all of the events are assumed to be unlensed. We can see a drastic improvement in the mis-classification when combined QT and SGP data representation (black curve) is used as opposed to using only single data representation of either QT (orange) or SGP (brown). }
    \label{fig:on_real_data}
\end{figure}

Next, we create larger test datasets of different kinds and compare their performances with each other along with results of G21 (see \Fref{fig:all_MQS}).
We evaluate the performance of our ML models on a test sample extracted from Dataset-G. Here, we take 2000 lensed pairs and 100,000 unlensed pairs. All of our models perform better than G21 even on this larger dataset (see {\it top left panel} of \Fref{fig:all_MQS}). Further, to compare between our own models, we find that Model Q$_{\rm r}$~S$_{\rm g}$ (solid black) performs better than Q$_{\rm r}$ (solid orange) demonstrating the importance of including constraints from SGP maps. Also, since the Model Q$_{\rm r}$~S$_{\rm g}$ (i.e. the network trained on QT maps from Dataset-R and SGP maps from Dataset-G) performs marginally better than Q$_{\rm r}$~S$_{\rm r}$. Hence, we choose Q$_{\rm r}$~S$_{\rm g}$ as our final best network model. We further find that the Model Q$_{\rm r}$ performs better on the QT maps test dataset made with Gaussian noise (dash-dotted orange) as compared to the real noise (solid orange).

Subsequently, we divide this dataset based on their signal-to-noise ratio (SNR), total mass ($M_{\rm t}$) and mass ratio ($q$, with $0 < q \leq 1$).  If the network SNR of both of the events in a pair is between 10 and 15 then its called high SNR while if the network SNRs of either of the events are $\leq$ 10, then this pair is considered as low SNR. High- and low- total mass (detector frame mass) correspond to $>$ 80~$M_\odot$ and $\leq$ 80~$M_\odot$, respectively. And, high- and low- mass ratio correspond to  $q>$ 0.6 and 0.3 $<q<~$0.6, respectively. Here, we compare performances of Models Q$_{\rm r}$~S$_{\rm g}$, Q$_{\rm r}$~S$_{\rm r}$ and Q$_{\rm r}$. 
From the \Fref{fig:all_MQS}, we can infer that the Model-Q$_{\rm r}$S$_{\rm g}$ (black) is almost always better than the other models, more so at lower FPRs.

As expected, our models perform better on high SNR (dashed) as compared to low SNR (dotted) shown in the {\it top right panel} of \Fref{fig:all_MQS}. In case of high- and low- $M_{\rm t}$, we find that our models perform better on high $M_{\rm t}$ as compared to low $M_{\rm t}$. And, in the case of high- and low- $q$, our models have better performance on low $q$ than on high $q$.


Lastly, we test our models on the real events detected by the LIGO--Virgo collaboration. Here, we use the Dataset-LHV (see \Tref{tab:datasets}) which contains 46 real events during O2 and O3. We pair each event with every other, this gives us a total of 1035 pairs. We evaluate the Model-Q$_{\rm r}$~S$_{\rm g}$ on this dataset assuming all of the 1035 event pairs are unlensed. As a metric, we plot the fraction of mis-classifications as a function of the network threshold which is the lower bound on the output probability of the network to consider an event pair to be lensed (see \Fref{fig:on_real_data}). 
Since the Model-Q$_{\rm r}$~S$_{\rm g}$ has two parallel networks one trained on the QT maps and the other on the SGP maps. In \Fref{fig:on_real_data}, we also show the fraction of mis-classifications for Model-Q$_{\rm r}$ (orange) and Model-S$_{\rm g}$ (brown) along with their combined output probability (black). This further demonstrates that the Model- Q$_{\rm r}$ or S$_{\rm g}$  alone has a higher number of mis-classifications whereas the combined Model Q$_{\rm r}$~S$_{\rm g}$ gives a drastically lower number of mis-classifications.

During our analysis, we noticed that superposition approach in the preparation of QT maps may be prone to producing more FPs. For instance, consider event 1 of the pair has a very high SNR and the event 2 has a low SNR (but still super-threshold) which are not lensed. The QT map of event 1 will show clear chirping signal, while the QT map of event 2 will be mostly noise-like because of the low SNR. When superposed, the resulting image is dominated by the stronger event 1. This mimics the appearance of a superposed lensed pair and will be classified as a lens even though the events are unlensed in reality. However, the 2x2 grid chosen for SGP maps is less likely to result in FPs as is evident from \Fref{fig:on_real_data}.

    
\section{Summary}
\label{sec:summ}
The current method to infer the probability for a pair of GW signals to be lensed is by comparing the posteriors of the intrinsic parameters. Lensed pairs will have large overlap in the posteriors while the unlensed pairs will not show significant overlap and the comparison is quantified using bayesian model selection referred to as the posterior overlap method. However, this technique is time consuming and can take hours to days for parameter estimation.
There are also some alternative pipelines- \texttt{GOLUM}, \texttt{HANABI} and \texttt{phazap}. However, in terms of computational speed,
these algorithms may take $\mathcal{O}(\rm hours)$ for an event pair. Further, in the future observing runs, such as O5 and voyager, the predicted number of detectable events are $\mathcal{O}(10^{3})-\mathcal{O}(10^{4})$, and for the 3G detectors, the predicted numbers are $\mathcal{O}(10^{6})$. Thus, the posterior overlap and other methods may be computationally expensive relative to machine learning algorithms. These pipelines could nevertheless serve as complementary techniques and could possibly be applied on the candidates selected by ML algorithm such as \texttt{SLICK}.

In \texttt{SLICK}, we use two neural network models - {\it Densenet121} and {\it 6 block CNN}, which are trained on the QT maps and SGP maps, respectively. The output from these two models is then combined to give a final output probability which tells whether the given pair of events is lensed or not. We work with 3 datasets - $1)$ GW signals injected in stationary gaussian O4 noise (Dataset-G), $2)$ GW signals injected in real O3 noise (Dataset-R) and 
$3)$ real GW events (Dataset-LHV). We use the first two datasets for training, validation and testing whereas the third dataset is for testing only. The networks are trained on different datasets to create 3 different ``best" models which are labelled as Model-Q$_{\rm r}$, Model-Q$_{\rm r}$S$_{\rm g}$ and Model-Q$_{\rm r}$S$_{\rm r}$ where Q and S refer to the type of data (QT or SGP maps, respectively) and the subscript r or g correspond to the training dataset based on the type of noise. 

Firstly, we test our models on the QT maps prepared from the same BBH sample as G21 (DST dataset) and compare the ROC curves. We find that Model-Q$_{\rm r}$ has an improved performance, especially, at lower FPRs ($<0.01$). We also infer that  Model-Q$_{\rm r}$S$_{\rm g}$ performs better than Model-Q$_{\rm r}$. Our inference does not change when we evaluate our models on our test datasets. Our models, when tested, on real GW events strengthens the confidence in Model-Q$_{\rm r}$S$_{\rm g}$ further as it significantly reduces misclassifications compared to models that use either QT or SGP data alone. Thus, the final best model of \texttt{SLICK} is chosen to be Model-Q$_{\rm r}$S$_{\rm g}$.

We note that sky localisation is expected to provide strong constraints for lensed event pairs (also, demonstrated by G21). Thus, including skymaps as additional type of input data in {\sc \texttt{SLICK}} will further improve its performance. However, it is not imperative to include it in the machine learning framework since skymaps can also be easily used as a screening step subsequently on the promising candidates. 

In the future, we plan to include signals with added complexities such as spinning/precessing and eccentric BBH signals. It will be interesting to see if \texttt{SLICK} can continue to give robust predictions with high efficiency allowing us correctly classify a broader sample of GW signals.  

\section*{Acknowledgements}
We would like to thank Soorya Narayan, Sudhagar S, Srashti Goyal and Shasvath Kapadia for useful discussions. We acknowledge the use of IUCAA LDG cluster Sarathi for the computational/numerical work. This material is based upon work supported by NSF's LIGO Laboratory which is a major facility fully funded by the National Science Foundation. This work uses \texttt{NumPy} \citep{harris2020array}, \texttt{SciPy} \citep{2020SciPy-NMeth}, \texttt{Pandas} \citep{mckinney-proc-scipy-2010}, \texttt{PyCBC} \citep{alex_nitz_2024_10473621}, \texttt{GWpy} \citep{gwpy}, \texttt{LALSuite} \citep{lalsuite}, \texttt{Tensorflow} \citep{tensorflow2015-whitepaper}
software packages. 
\section*{Data Availability}
The \texttt{SLICK} pipeline and the data used in the training and testing will be available upon reasonable request to the authors.
 



\bibliographystyle{mnras}
\bibliography{lensing} 

\begin{thebibliography}{}
\makeatletter
\relax
\def\mn@urlcharsother{\let\do\@makeother \do\$\do\&\do\#\do\^\do\_\do\%\do\~}
\def\mn@doi{\begingroup\mn@urlcharsother \@ifnextchar [ {\mn@doi@} {\mn@doi@[]}}
\def\mn@doi@[#1]#2{\def\@tempa{#1}\ifx\@tempa\@empty \href {http://dx.doi.org/#2} {doi:#2}\else \href {http://dx.doi.org/#2} {#1}\fi \endgroup}
\def\mn@eprint#1#2{\mn@eprint@#1:#2::\@nil}
\def\mn@eprint@arXiv#1{\href {http://arxiv.org/abs/#1} {{\tt arXiv:#1}}}
\def\mn@eprint@dblp#1{\href {http://dblp.uni-trier.de/rec/bibtex/#1.xml} {dblp:#1}}
\def\mn@eprint@#1:#2:#3:#4\@nil{\def\@tempa {#1}\def\@tempb {#2}\def\@tempc {#3}\ifx \@tempc \@empty \let \@tempc \@tempb \let \@tempb \@tempa \fi \ifx \@tempb \@empty \def\@tempb {arXiv}\fi \@ifundefined {mn@eprint@\@tempb}{\@tempb:\@tempc}{\expandafter \expandafter \csname mn@eprint@\@tempb\endcsname \expandafter{\@tempc}}}

\bibitem[\protect\citeauthoryear{Abadi et~al.,}{Abadi et~al.}{2015}]{tensorflow2015-whitepaper}
Abadi M.,  et~al., 2015, {TensorFlow}: Large-Scale Machine Learning on Heterogeneous Systems, \url {http://tensorflow.org/}

\bibitem[\protect\citeauthoryear{Abbott et~al.,}{Abbott et~al.}{2016}]{PhysRevLett.116.061102}
Abbott B.~P.,  et~al., 2016, \mn@doi [Phys. Rev. Lett.] {10.1103/PhysRevLett.116.061102}, 116, 061102

\bibitem[\protect\citeauthoryear{Abbott et~al.,}{Abbott et~al.}{2017a}]{PhysRevLett.119.161101}
Abbott B.~P.,  et~al., 2017a, \mn@doi [Phys. Rev. Lett.] {10.1103/PhysRevLett.119.161101}, 119, 161101

\bibitem[\protect\citeauthoryear{Abbott et~al.,}{Abbott et~al.}{2017b}]{Abbott_2017}
Abbott B.~P.,  et~al., 2017b, \mn@doi [The Astrophysical Journal Letters] {10.3847/2041-8213/aa91c9}, 848, L12

\bibitem[\protect\citeauthoryear{Abbott et~al.,}{Abbott et~al.}{2017c}]{Abbott_2017b}
Abbott B.~P.,  et~al., 2017c, \mn@doi [The Astrophysical Journal Letters] {10.3847/2041-8213/aa920c}, 848, L13

\bibitem[\protect\citeauthoryear{Abbott et~al.,}{Abbott et~al.}{2019a}]{PhysRevX.9.031040}
Abbott B.~P.,  et~al., 2019a, \mn@doi [Phys. Rev. X] {10.1103/PhysRevX.9.031040}, 9, 031040

\bibitem[\protect\citeauthoryear{Abbott et~al.,}{Abbott et~al.}{2019b}]{PhysRevD.100.104036}
Abbott B.~P.,  et~al., 2019b, \mn@doi [Phys. Rev. D] {10.1103/PhysRevD.100.104036}, 100, 104036

\bibitem[\protect\citeauthoryear{Abbott et~al.,}{Abbott et~al.}{2021a}]{PhysRevX.11.021053}
Abbott R.,  et~al., 2021a, \mn@doi [Phys. Rev. X] {10.1103/PhysRevX.11.021053}, 11, 021053

\bibitem[\protect\citeauthoryear{Abbott et~al.,}{Abbott et~al.}{2021b}]{Abbott_2021}
Abbott R.,  et~al., 2021b, \mn@doi [Physical Review X] {10.1103/physrevx.11.021053}, 11

\bibitem[\protect\citeauthoryear{Abbott et~al.}{Abbott et~al.}{2021c}]{LIGOScientific:2019lzm}
Abbott R.,  et~al., 2021c, \mn@doi [SoftwareX] {10.1016/j.softx.2021.100658}, 13, 100658

\bibitem[\protect\citeauthoryear{Abbott et~al.,}{Abbott et~al.}{2021d}]{Abbott_2021_pop}
Abbott R.,  et~al., 2021d, \mn@doi [The Astrophysical Journal Letters] {10.3847/2041-8213/abe949}, 913, L7

\bibitem[\protect\citeauthoryear{Abbott et~al.,}{Abbott et~al.}{2021e}]{Abbott_2021_lnsing}
Abbott R.,  et~al., 2021e, \mn@doi [The Astrophysical Journal] {10.3847/1538-4357/ac23db}, 923, 14

\bibitem[\protect\citeauthoryear{Abbott et~al.,}{Abbott et~al.}{2023a}]{PhysRevX.13.041039}
Abbott R.,  et~al., 2023a, \mn@doi [Phys. Rev. X] {10.1103/PhysRevX.13.041039}, 13, 041039

\bibitem[\protect\citeauthoryear{Abbott et~al.}{Abbott et~al.}{2023b}]{KAGRA:2023pio}
Abbott R.,  et~al., 2023b, \mn@doi [Astrophys. J. Suppl.] {10.3847/1538-4365/acdc9f}, 267, 29

\bibitem[\protect\citeauthoryear{Bahaadini, Rohani, Coughlin, Zevin, Kalogera  \& Katsaggelos}{Bahaadini et~al.}{2017}]{bahaadini2017deep}
Bahaadini S.,  Rohani N.,  Coughlin S.,  Zevin M.,  Kalogera V.,   Katsaggelos A.~K.,  2017, Deep Multi-view Models for Glitch Classification (\mn@eprint {arXiv} {1705.00034})

\bibitem[\protect\citeauthoryear{Basak, Ganguly, Haris, Kapadia, Mehta  \& Ajith}{Basak et~al.}{2022}]{Basak_2022}
Basak S.,  Ganguly A.,  Haris K.,  Kapadia S.,  Mehta A.~K.,   Ajith P.,  2022, \mn@doi [The Astrophysical Journal Letters] {10.3847/2041-8213/ac4dfa}, 926, L28

\bibitem[\protect\citeauthoryear{Cavaglia, Staats  \& Gill}{Cavaglia et~al.}{2019}]{Cavaglia_2019}
Cavaglia M.,  Staats K.,   Gill T.,  2019, \mn@doi [Communications in Computational Physics] {10.4208/cicp.oa-2018-0092}, 25

\bibitem[\protect\citeauthoryear{Chatterji, Blackburn, Martin  \& Katsavounidis}{Chatterji et~al.}{2004}]{SChatterji_2004}
Chatterji S.,  Blackburn L.,  Martin G.,   Katsavounidis E.,  2004, \mn@doi [Classical and Quantum Gravity] {10.1088/0264-9381/21/20/024}, 21, S1809

\bibitem[\protect\citeauthoryear{Cheung, Rinaldi, Toscani  \& Hannuksela}{Cheung et~al.}{2023}]{cheung2023mitigating}
Cheung D. H.~T.,  Rinaldi S.,  Toscani M.,   Hannuksela O.~A.,  2023, Mitigating the effect of population model uncertainty on strong lensing Bayes factor using nonparametric methods (\mn@eprint {arXiv} {2308.12182})

\bibitem[\protect\citeauthoryear{Choudhary, More, Suyamprakasam  \& Bose}{Choudhary et~al.}{2023}]{PhysRevD.107.024030}
Choudhary S.,  More A.,  Suyamprakasam S.,   Bose S.,  2023, \mn@doi [Phys. Rev. D] {10.1103/PhysRevD.107.024030}, 107, 024030

\bibitem[\protect\citeauthoryear{Collaboration et~al.,}{Collaboration et~al.}{2023}]{theligoscientificcollaboration2023search}
Collaboration T. L.~S.,  et~al., 2023, Search for gravitational-lensing signatures in the full third observing run of the LIGO-Virgo network (\mn@eprint {arXiv} {2304.08393})

\bibitem[\protect\citeauthoryear{{Dai} \& {Venumadhav}}{{Dai} \& {Venumadhav}}{2017}]{Dai2017}
{Dai} L.,  {Venumadhav} T.,  2017, arXiv e-prints, \href {https://ui.adsabs.harvard.edu/abs/2017arXiv170204724D} {p. arXiv:1702.04724}

\bibitem[\protect\citeauthoryear{{Dai}, {Zackay}, {Venumadhav}, {Roulet}  \& {Zaldarriaga}}{{Dai} et~al.}{2020}]{Dai2020Morse}
{Dai} L.,  {Zackay} B.,  {Venumadhav} T.,  {Roulet} J.,   {Zaldarriaga} M.,  2020, arXiv e-prints, \href {https://ui.adsabs.harvard.edu/abs/2020arXiv200712709D} {p. arXiv:2007.12709}

\bibitem[\protect\citeauthoryear{{Deguchi} \& {Watson}}{{Deguchi} \& {Watson}}{1986}]{1986ApJ...307...30D}
{Deguchi} S.,  {Watson} W.~D.,  1986, \mn@doi [\apj] {10.1086/164389}, \href {https://ui.adsabs.harvard.edu/abs/1986ApJ...307...30D} {307, 30}

\bibitem[\protect\citeauthoryear{Deng, Dong, Socher, Li, Li  \& Fei-Fei}{Deng et~al.}{2009}]{imgNetdata}
Deng J.,  Dong W.,  Socher R.,  Li L.-J.,  Li K.,   Fei-Fei L.,  2009, in 2009 IEEE Conference on Computer Vision and Pattern Recognition. pp 248--255, \mn@doi{10.1109/CVPR.2009.5206848}

\bibitem[\protect\citeauthoryear{Dodelson}{Dodelson}{2017}]{dodelson2017}
Dodelson S.,  2017, Gravitational lensing.
Cambridge University Press

\bibitem[\protect\citeauthoryear{Ezquiaga, Hu  \& Lo}{Ezquiaga et~al.}{2023}]{Ezquiaga:2023xfe}
Ezquiaga J.~M.,  Hu W.,   Lo R. K.~L.,  2023, \mn@doi [Phys. Rev. D] {10.1103/PhysRevD.108.103520}, 108, 103520

\bibitem[\protect\citeauthoryear{Fan, Liao, Biesiada, Pi\'orkowska-Kurpas  \& Zhu}{Fan et~al.}{2017}]{PhysRevLett.118.091102}
Fan X.-L.,  Liao K.,  Biesiada M.,  Pi\'orkowska-Kurpas A.,   Zhu Z.-H.,  2017, \mn@doi [Phys. Rev. Lett.] {10.1103/PhysRevLett.118.091102}, 118, 091102

\bibitem[\protect\citeauthoryear{George, Shen  \& Huerta}{George et~al.}{2018}]{PhysRevD.97.101501}
George D.,  Shen H.,   Huerta E.~A.,  2018, \mn@doi [Phys. Rev. D] {10.1103/PhysRevD.97.101501}, 97, 101501

\bibitem[\protect\citeauthoryear{Gong, Hou, Liang  \& Papantonopoulos}{Gong et~al.}{2018}]{PhysRevD.97.084040}
Gong Y.,  Hou S.,  Liang D.,   Papantonopoulos E.,  2018, \mn@doi [Phys. Rev. D] {10.1103/PhysRevD.97.084040}, 97, 084040

\bibitem[\protect\citeauthoryear{Goyal, Haris, Mehta  \& Ajith}{Goyal et~al.}{2021a}]{PhysRevD.103.024038}
Goyal S.,  Haris K.,  Mehta A.~K.,   Ajith P.,  2021a, \mn@doi [Phys. Rev. D] {10.1103/PhysRevD.103.024038}, 103, 024038

\bibitem[\protect\citeauthoryear{Goyal, D., Kapadia  \& Ajith}{Goyal et~al.}{2021b}]{PhysRevD.104.124057}
Goyal S.,  D. H.,  Kapadia S.~J.,   Ajith P.,  2021b, \mn@doi [Phys. Rev. D] {10.1103/PhysRevD.104.124057}, 104, 124057

\bibitem[\protect\citeauthoryear{{Gunn}}{{Gunn}}{1967}]{gunn1967}
{Gunn} J.~E.,  1967, \mn@doi [\apj] {10.1086/149378}, \href {https://ui.adsabs.harvard.edu/abs/1967ApJ...150..737G} {150, 737}

\bibitem[\protect\citeauthoryear{Hannam, Schmidt, Boh\'e, Haegel, Husa, Ohme, Pratten  \& P\"urrer}{Hannam et~al.}{2014}]{PhysRevLett.113.151101}
Hannam M.,  Schmidt P.,  Boh\'e A.,  Haegel L.,  Husa S.,  Ohme F.,  Pratten G.,   P\"urrer M.,  2014, \mn@doi [Phys. Rev. Lett.] {10.1103/PhysRevLett.113.151101}, 113, 151101

\bibitem[\protect\citeauthoryear{{Hannuksela}, {Haris}, {Ng}, {Kumar}, {Mehta}, {Keitel}, {Li}  \& {Ajith}}{{Hannuksela} et~al.}{2019}]{Hannuksela2019}
{Hannuksela} O.~A.,  {Haris} K.,  {Ng} K.~K.~Y.,  {Kumar} S.,  {Mehta} A.~K.,  {Keitel} D.,  {Li} T.~G.~F.,   {Ajith} P.,  2019, \mn@doi [\apjl] {10.3847/2041-8213/ab0c0f}, \href {https://ui.adsabs.harvard.edu/abs/2019ApJ...874L...2H} {874, L2}

\bibitem[\protect\citeauthoryear{{Haris}, {Mehta}, {Kumar}, {Venumadhav}  \& {Ajith}}{{Haris} et~al.}{2018}]{Haris2018}
{Haris} K.,  {Mehta} A.~K.,  {Kumar} S.,  {Venumadhav} T.,   {Ajith} P.,  2018, arXiv e-prints, \href {https://ui.adsabs.harvard.edu/abs/2018arXiv180707062H} {p. arXiv:1807.07062}

\bibitem[\protect\citeauthoryear{Harris et~al.,}{Harris et~al.}{2020}]{harris2020array}
Harris C.~R.,  et~al., 2020, \mn@doi [Nature] {10.1038/s41586-020-2649-2}, 585, 357

\bibitem[\protect\citeauthoryear{Huang, Liu, van~der Maaten  \& Weinberger}{Huang et~al.}{2018}]{huang2018densely}
Huang G.,  Liu Z.,  van~der Maaten L.,   Weinberger K.~Q.,  2018, Densely Connected Convolutional Networks (\mn@eprint {arXiv} {1608.06993})

\bibitem[\protect\citeauthoryear{Jacobs, Glazebrook, Collett, More  \& McCarthy}{Jacobs et~al.}{2017}]{10.1093/mnras/stx1492}
Jacobs C.,  Glazebrook K.,  Collett T.,  More A.,   McCarthy C.,  2017, \mn@doi [Monthly Notices of the Royal Astronomical Society] {10.1093/mnras/stx1492}, 471, 167

\bibitem[\protect\citeauthoryear{Jacobs et~al.,}{Jacobs et~al.}{2019}]{Jacobs_2019}
Jacobs C.,  et~al., 2019, \mn@doi [The Astrophysical Journal Supplement Series] {10.3847/1538-4365/ab26b6}, 243, 17

\bibitem[\protect\citeauthoryear{Jaelani, More, Wong, Inoue, Chao, Premadi  \& Cañameras}{Jaelani et~al.}{2023}]{jaelani2023survey}
Jaelani A.~T.,  More A.,  Wong K.~C.,  Inoue K.~T.,  Chao D. C.~Y.,  Premadi P.~W.,   Cañameras R.,  2023, Survey of Gravitationally lensed Objects in HSC Imaging (SuGOHI) $-$ X. Strong Lens Finding in The HSC-SSP using Convolutional Neural Networks (\mn@eprint {arXiv} {2312.07333})

\bibitem[\protect\citeauthoryear{Jana, Kapadia, Venumadhav  \& Ajith}{Jana et~al.}{2023}]{PhysRevLett.130.261401}
Jana S.,  Kapadia S.~J.,  Venumadhav T.,   Ajith P.,  2023, \mn@doi [Phys. Rev. Lett.] {10.1103/PhysRevLett.130.261401}, 130, 261401

\bibitem[\protect\citeauthoryear{Janquart, Hannuksela, Haris  \& Broeck}{Janquart et~al.}{2022a}]{janquart2022golum}
Janquart J.,  Hannuksela O.~A.,  Haris K.,   Broeck C. V.~D.,  2022a, GOLUM: A fast and precise methodology to search for, and analyze, strongly lensed gravitational-wave events (\mn@eprint {arXiv} {2203.06444})

\bibitem[\protect\citeauthoryear{Janquart, More  \& Van Den~Broeck}{Janquart et~al.}{2022b}]{Janquart:2022zdd}
Janquart J.,  More A.,   Van Den~Broeck C.,  2022b, \mn@doi [Mon. Not. Roy. Astron. Soc.] {10.1093/mnras/stac3660}, 519, 2046

\bibitem[\protect\citeauthoryear{Janquart et~al.,}{Janquart et~al.}{2023}]{10.1093/mnras/stad2909}
Janquart J.,  et~al., 2023, \mn@doi [Monthly Notices of the Royal Astronomical Society] {10.1093/mnras/stad2909}, 526, 3832

\bibitem[\protect\citeauthoryear{Khan, Husa, Hannam, Ohme, P\"urrer, Forteza  \& Boh\'e}{Khan et~al.}{2016}]{PhysRevD.93.044007}
Khan S.,  Husa S.,  Hannam M.,  Ohme F.,  P\"urrer M.,  Forteza X.~J.,   Boh\'e A.,  2016, \mn@doi [Phys. Rev. D] {10.1103/PhysRevD.93.044007}, 93, 044007

\bibitem[\protect\citeauthoryear{{Kormann}, {Schneider}  \& {Bartelmann}}{{Kormann} et~al.}{1994}]{Kormann1994}
{Kormann} R.,  {Schneider} P.,   {Bartelmann} M.,  1994, \aap, \href {https://ui.adsabs.harvard.edu/abs/1994A&A...284..285K} {284, 285}

\bibitem[\protect\citeauthoryear{Krizhevsky, Sutskever  \& Hinton}{Krizhevsky et~al.}{2012}]{ImgNetClass}
Krizhevsky A.,  Sutskever I.,   Hinton G.,  2012, \mn@doi [Neural Information Processing Systems] {10.1145/3065386}, 25

\bibitem[\protect\citeauthoryear{{LIGO Scientific Collaboration}, {Virgo Collaboration}  \& {KAGRA Collaboration}}{{LIGO Scientific Collaboration} et~al.}{2018}]{lalsuite}
{LIGO Scientific Collaboration} {Virgo Collaboration}  {KAGRA Collaboration} 2018, {LVK} {A}lgorithm {L}ibrary - {LALS}uite, Free software (GPL), \mn@doi{10.7935/GT1W-FZ16}

\bibitem[\protect\citeauthoryear{Lai, Hannuksela, Herrera-Mart\'{\i}n, Diego, Broadhurst  \& Li}{Lai et~al.}{2018}]{PhysRevD.98.083005}
Lai K.-H.,  Hannuksela O.~A.,  Herrera-Mart\'{\i}n A.,  Diego J.~M.,  Broadhurst T.,   Li T. G.~F.,  2018, \mn@doi [Phys. Rev. D] {10.1103/PhysRevD.98.083005}, 98, 083005

\bibitem[\protect\citeauthoryear{LeCun, Bengio  \& Hinton}{LeCun et~al.}{2015}]{deep_article}
LeCun Y.,  Bengio Y.,   Hinton G.,  2015, \mn@doi [Nature] {10.1038/nature14539}, 521, 436

\bibitem[\protect\citeauthoryear{Li, Mao, Zhao  \& Lu}{Li et~al.}{2018}]{10.1093/mnras/sty411}
Li S.-S.,  Mao S.,  Zhao Y.,   Lu Y.,  2018, \mn@doi [Monthly Notices of the Royal Astronomical Society] {10.1093/mnras/sty411}, 476, 2220

\bibitem[\protect\citeauthoryear{Li, Fan  \& Gou}{Li et~al.}{2019}]{Li_2019}
Li Y.,  Fan X.,   Gou L.,  2019, \mn@doi [The Astrophysical Journal] {10.3847/1538-4357/ab037e}, 873, 37

\bibitem[\protect\citeauthoryear{{Li} et~al.,}{{Li} et~al.}{2021}]{2021ApJ...923...16L}
{Li} R.,  et~al., 2021, \mn@doi [\apj] {10.3847/1538-4357/ac2df0}, \href {https://ui.adsabs.harvard.edu/abs/2021ApJ...923...16L} {923, 16}

\bibitem[\protect\citeauthoryear{Lo \& Magaña~Hernandez}{Lo \& Magaña~Hernandez}{2023}]{Lo_2023}
Lo R.~K.,  Magaña~Hernandez I.,  2023, \mn@doi [Physical Review D] {10.1103/physrevd.107.123015}, 107

\bibitem[\protect\citeauthoryear{{Macleod}, {Areeda}, {Coughlin}, {Massinger}  \& {Urban}}{{Macleod} et~al.}{2021}]{gwpy}
{Macleod} D.~M.,  {Areeda} J.~S.,  {Coughlin} S.~B.,  {Massinger} T.~J.,   {Urban} A.~L.,  2021, \mn@doi [SoftwareX] {10.1016/j.softx.2021.100657}, 13, 100657

\bibitem[\protect\citeauthoryear{Magare, Kapadia, More, Singh, Ajith  \& Ramprakash}{Magare et~al.}{2023}]{Magare_2023}
Magare S.,  Kapadia S.~J.,  More A.,  Singh M.~K.,  Ajith P.,   Ramprakash A.~N.,  2023, \mn@doi [The Astrophysical Journal Letters] {10.3847/2041-8213/acf668}, 955, L31

\bibitem[\protect\citeauthoryear{Maggiore et~al.,}{Maggiore et~al.}{2020}]{Maggiore_2020et}
Maggiore M.,  et~al., 2020, \mn@doi [Journal of Cosmology and Astroparticle Physics] {10.1088/1475-7516/2020/03/050}, 2020, 050

\bibitem[\protect\citeauthoryear{{Miller} \& {Yunes}}{{Miller} \& {Yunes}}{2019}]{2019Natur.568..469M}
{Miller} M.~C.,  {Yunes} N.,  2019, \mn@doi [\nat] {10.1038/s41586-019-1129-z}, \href {https://ui.adsabs.harvard.edu/abs/2019Natur.568..469M} {568, 469}

\bibitem[\protect\citeauthoryear{More \& More}{More \& More}{2022}]{10.1093/mnras/stac1704}
More A.,  More S.,  2022, \mn@doi [Monthly Notices of the Royal Astronomical Society] {10.1093/mnras/stac1704}, 515, 1044

\bibitem[\protect\citeauthoryear{Nakamura}{Nakamura}{1998}]{PhysRevLett.80.1138}
Nakamura T.~T.,  1998, \mn@doi [Phys. Rev. Lett.] {10.1103/PhysRevLett.80.1138}, 80, 1138

\bibitem[\protect\citeauthoryear{Ng, Wong, Broadhurst  \& Li}{Ng et~al.}{2018}]{PhysRevD.97.023012}
Ng K. K.~Y.,  Wong K. W.~K.,  Broadhurst T.,   Li T. G.~F.,  2018, \mn@doi [Phys. Rev. D] {10.1103/PhysRevD.97.023012}, 97, 023012

\bibitem[\protect\citeauthoryear{Nitz et~al.,}{Nitz et~al.}{2024}]{alex_nitz_2024_10473621}
Nitz A.,  et~al., 2024, gwastro/pycbc: v2.3.3 release of PyCBC, \mn@doi{10.5281/zenodo.10473621}, \url {https://doi.org/10.5281/zenodo.10473621}

\bibitem[\protect\citeauthoryear{Oguri}{Oguri}{2018}]{10.1093/mnras/sty2145}
Oguri M.,  2018, \mn@doi [Monthly Notices of the Royal Astronomical Society] {10.1093/mnras/sty2145}, 480, 3842

\bibitem[\protect\citeauthoryear{Oguri \& Takahashi}{Oguri \& Takahashi}{2020}]{Oguri_2020}
Oguri M.,  Takahashi R.,  2020, \mn@doi [The Astrophysical Journal] {10.3847/1538-4357/abafab}, 901, 58

\bibitem[\protect\citeauthoryear{Oost, Mukohyama  \& Wang}{Oost et~al.}{2018}]{PhysRevD.97.124023}
Oost J.,  Mukohyama S.,   Wang A.,  2018, \mn@doi [Phys. Rev. D] {10.1103/PhysRevD.97.124023}, 97, 124023

\bibitem[\protect\citeauthoryear{Reitze et~al.,}{Reitze et~al.}{2019}]{Reitze2019Cosmic}
Reitze D.,  et~al., 2019, Bulletin of the AAS, 51

\bibitem[\protect\citeauthoryear{Schmidhuber}{Schmidhuber}{2015}]{SCHMIDHUBER201585}
Schmidhuber J.,  2015, \mn@doi [Neural Networks] {https://doi.org/10.1016/j.neunet.2014.09.003}, 61, 85

\bibitem[\protect\citeauthoryear{{Schneider}, {Ehlers}  \& {Falco}}{{Schneider} et~al.}{1992}]{schneider1992}
{Schneider} P.,  {Ehlers} J.,   {Falco} E.~E.,  1992, {Gravitational Lenses}.
Springer, Berlin, Heidelberg

\bibitem[\protect\citeauthoryear{Schutz}{Schutz}{2009}]{schutz_2009}
Schutz B.,  2009, A First Course in General Relativity, 2 edn.
Cambridge University Press, \mn@doi{10.1017/CBO9780511984181}

\bibitem[\protect\citeauthoryear{Sereno, Jetzer, Sesana  \& Volonteri}{Sereno et~al.}{2011}]{10.1111/j.1365-2966.2011.18895.x}
Sereno M.,  Jetzer P.,  Sesana A.,   Volonteri M.,  2011, \mn@doi [Monthly Notices of the Royal Astronomical Society] {10.1111/j.1365-2966.2011.18895.x}, 415, 2773

\bibitem[\protect\citeauthoryear{Shen, George, Huerta  \& Zhao}{Shen et~al.}{2019}]{Shen_2019}
Shen H.,  George D.,  Huerta E.~A.,   Zhao Z.,  2019, in {ICASSP} 2019 - 2019 {IEEE} International Conference on Acoustics, Speech and Signal Processing ({ICASSP}). {IEEE}, \mn@doi{10.1109/icassp.2019.8683061}, \url {https://doi.org/10.1109%2Ficassp.2019.8683061}

\bibitem[\protect\citeauthoryear{Singer \& Price}{Singer \& Price}{2016}]{PhysRevD.93.024013}
Singer L.~P.,  Price L.~R.,  2016, \mn@doi [Phys. Rev. D] {10.1103/PhysRevD.93.024013}, 93, 024013

\bibitem[\protect\citeauthoryear{Singer, Goldstein  \& Bloom}{Singer et~al.}{2019}]{singer2019ligovirgo}
Singer L.~P.,  Goldstein D.~A.,   Bloom J.~S.,  2019, The Two LIGO/Virgo Binary Black Hole Mergers on 2019 August 28 Were Not Strongly Lensed (\mn@eprint {arXiv} {1910.03601})

\bibitem[\protect\citeauthoryear{Smith, Field, Blackburn, Haster, P\"urrer, Raymond  \& Schmidt}{Smith et~al.}{2016}]{PhysRevD.94.044031}
Smith R.,  Field S.~E.,  Blackburn K.,  Haster C.-J.,  P\"urrer M.,  Raymond V.,   Schmidt P.,  2016, \mn@doi [Phys. Rev. D] {10.1103/PhysRevD.94.044031}, 94, 044031

\bibitem[\protect\citeauthoryear{Takahashi \& Nakamura}{Takahashi \& Nakamura}{2003}]{Takahashi_2003}
Takahashi R.,  Nakamura T.,  2003, \mn@doi [The Astrophysical Journal] {10.1086/377430}, 595, 1039

\bibitem[\protect\citeauthoryear{Virtanen et~al.,}{Virtanen et~al.}{2020}]{2020SciPy-NMeth}
Virtanen P.,  et~al., 2020, \mn@doi [Nature Methods] {10.1038/s41592-019-0686-2}, \href {https://rdcu.be/b08Wh} {17, 261}

\bibitem[\protect\citeauthoryear{{Wang}, {Stebbins}  \& {Turner}}{{Wang} et~al.}{1996}]{Wang1996}
{Wang} Y.,  {Stebbins} A.,   {Turner} E.~L.,  1996, \mn@doi [\prl] {10.1103/PhysRevLett.77.2875}, \href {https://ui.adsabs.harvard.edu/abs/1996PhRvL..77.2875W} {77, 2875}

\bibitem[\protect\citeauthoryear{Wei et~al.,}{Wei et~al.}{2017}]{Wei_2017}
Wei J.-J.,  et~al., 2017, \mn@doi [Journal of Cosmology and Astroparticle Physics] {10.1088/1475-7516/2017/11/035}, 2017, 035

\bibitem[\protect\citeauthoryear{{W}es {M}c{K}inney}{{W}es {M}c{K}inney}{2010}]{mckinney-proc-scipy-2010}
{W}es {M}c{K}inney 2010, in {S}t\'efan van~der {W}alt {J}arrod {M}illman eds, {P}roceedings of the 9th {P}ython in {S}cience {C}onference. pp 56 -- 61, \mn@doi{10.25080/Majora-92bf1922-00a}

\bibitem[\protect\citeauthoryear{Xu, Ezquiaga  \& Holz}{Xu et~al.}{2022}]{Xu_2022}
Xu F.,  Ezquiaga J.~M.,   Holz D.~E.,  2022, \mn@doi [The Astrophysical Journal] {10.3847/1538-4357/ac58f8}, 929, 9

\bibitem[\protect\citeauthoryear{Yosinski, Clune, Bengio  \& Lipson}{Yosinski et~al.}{2014}]{yosinski2014transferable}
Yosinski J.,  Clune J.,  Bengio Y.,   Lipson H.,  2014, How transferable are features in deep neural networks? (\mn@eprint {arXiv} {1411.1792})

\bibitem[\protect\citeauthoryear{{Zaborowski} et~al.,}{{Zaborowski} et~al.}{2023}]{2023ApJ...954...68Z}
{Zaborowski} E.~A.,  et~al., 2023, \mn@doi [\apj] {10.3847/1538-4357/ace4ba}, \href {https://ui.adsabs.harvard.edu/abs/2023ApJ...954...68Z} {954, 68}

\bibitem[\protect\citeauthoryear{\c{C}al\i{}\c{s}kan, Ji, Cotesta, Berti, Kamionkowski  \& Marsat}{\c{C}al\i{}\c{s}kan et~al.}{2023a}]{Caliskan:2022hbu}
\c{C}al\i{}\c{s}kan M.,  Ji L.,  Cotesta R.,  Berti E.,  Kamionkowski M.,   Marsat S.,  2023a, \mn@doi [Phys. Rev. D] {10.1103/PhysRevD.107.043029}, 107, 043029

\bibitem[\protect\citeauthoryear{\c{C}al\i{}\c{s}kan, Ezquiaga, Hannuksela  \& Holz}{\c{C}al\i{}\c{s}kan et~al.}{2023b}]{Caliskan:2022wbh}
\c{C}al\i{}\c{s}kan M.,  Ezquiaga J.~M.,  Hannuksela O.~A.,   Holz D.~E.,  2023b, \mn@doi [Phys. Rev. D] {10.1103/PhysRevD.107.063023}, 107, 063023

\bibitem[\protect\citeauthoryear{\c{C}al\i{}\c{s}kan, Anil~Kumar, Ji, Ezquiaga, Cotesta, Berti  \& Kamionkowski}{\c{C}al\i{}\c{s}kan et~al.}{2023c}]{Caliskan:2023zqm}
\c{C}al\i{}\c{s}kan M.,  Anil~Kumar N.,  Ji L.,  Ezquiaga J.~M.,  Cotesta R.,  Berti E.,   Kamionkowski M.,  2023c, \mn@doi [Phys. Rev. D] {10.1103/PhysRevD.108.123543}, 108, 123543

\makeatother
\end{thebibliography}








\bsp	
\label{lastpage}
\end{document}